\documentclass{article}
\usepackage{authblk}
\usepackage{graphicx}
\usepackage{amsmath}
\usepackage{amssymb}
\usepackage{hyperref}
\usepackage{cleveref}
\usepackage{multirow,color}
\usepackage[export]{adjustbox}

\usepackage{tikz}
\usepackage{tikzit}

\tikzstyle{red node}=[fill=red, tikzit category=nodes, shape=circle, draw=black]
\tikzstyle{blue node}=[fill=blue, shape=circle, draw=black, tikzit category=nodes]
\tikzstyle{green node}=[tikzit fill=green, fill=green, shape=circle, draw=black, tikzit category=nodes]
\tikzstyle{yellow square}=[draw=black, fill=yellow, shape=rectangle]
\tikzstyle{blue node 2}=[fill={rgb,255: red,128; green,0; blue,128}, draw=black, shape=circle, tikzit fill=blue]
\tikzstyle{dasheda arrow}=[dashed, -stealth, line width=0.15mm]
\tikzstyle{dashedArrow}=[-latex, dashed, -stealth]
\tikzstyle{HeadStyle}=[>=latex]
\tikzstyle{DiscVert}=[fill=white,rotate=90,inner sep=-2.5pt,outer sep=0]
\tikzstyle{DiscHor}=[fill=white,rotate=0,inner sep=-2.5pt,outer sep=0]

\tikzstyle{dashed edge}=[dashed, -, line width=0.4]
\tikzstyle{blue pointer}=[->, draw=blue]
\tikzstyle{dashed thin arrow}=[->, line width=0.5, dashed, >=latex]
\tikzstyle{triangle}=[-, draw=none]
\tikzstyle{triagle}=[-, draw=triangle]
\tikzstyle{Line thick}=[-, line width=2.5, draw=black]
\tikzstyle{thin}=[-, line width=0.4]
\tikzstyle{thin Arrow}=[<->, line width=0.5, solid, >=latex]
\tikzstyle{dotdashed}=[-, dash dot, line width=0.5]
\tikzstyle{solid thin arrow}=[->, line width=0.5, solid, >=latex]
\tikzstyle{solid thick arrow}=[->, line width=2.5, solid, >=latex]
\tikzstyle{solid mid arrow}=[->, line width=1, solid, >=latex]

\input{Figures/Diagrams/sample.tikzdefs}
\usetikzlibrary{matrix,positioning,decorations.pathreplacing}

\usepackage{subfig}

\newcommand{\beqn}{\begin{equation}}
\newcommand{\eeqn}{\end{equation}}

\newcommand{\be}[1]{ \begin{equation} \label{eq:#1}}
\newcommand{\ee}{\end{equation}}
\newcommand{\bes}[1]{ \begin{equation} \label{eq:#1}\begin{array}{rl}}
\newcommand{\ees}{\end{array}\end{equation}}

\newcommand{\besp}{ \begin{split} }
\newcommand{\eesp}{\end{split}}

\providecommand\Rey{\mbox{\textit{Re}}}

\newcommand\im{\mathrm{i}}

\newcommand{\Ma}[1]{M}
\newcommand{\Pra}[1]{Pr}

\newcommand{\Bop}{\vec{B}}
\newcommand{\Fop}{\vec{F}}
\newcommand{\Lop}{\vec{L}}
\newcommand{\Gop}{\vec{G}}
\newcommand{\Nop}{\vec{N}}

\newcommand{\QV}{\vec{Q}} 
\newcommand{\qMV}{ \hat{\vec{q}} } 
\newcommand{\QVBF}{\vec{Q}_0} 
\newcommand{\parV}{\bm{\eta}} 

\newcommand{\qMVZ}[1]{ \qMV_{(z_{#1})}  }

\let\vec\mathbf

\usepackage{soul}
\usepackage{cancel}
\setstcolor{red}

\title{Mode selection in concentric jets. The steady-steady 1:2 resonant mode interaction with O(2) symmetry. }
\date{}
\author[1]{A. Corrochano}
\author[2,3]{J. Sierra-Ausín}
\author[1]{J. A. Martin}
\author[2]{\hspace{5cm}D. Fabre}
\author[1]{S. Le Clainche \thanks{Email address for correspondence: soledad.leclainche@upm.es}}

\affil[1]{School of Aerospace Engineering, Universidad Polit{\'e}cnica de Madrid, Madrid 28040, Spain}
\affil[2]{Institut de M{\'e}canique des Fluides de Toulouse (IMFT), Toulouse 31400, France}
\affil[3]{DIIN, Universit´a degli Studi di Salerno, Via Giovanni Paolo II, 84084 Fisciano (SA), Italy}

\begin{document}

\maketitle

\begin{abstract}
In this article, a thorough characterization of the configuration composed by two concentric jets at a low Reynolds number is presented. The analysis comprises a layout with a wide range for the velocity ratio between the inner and outer jets, defined within the interval $[0,2]$, and also details the influence of the distance between jets, where the wall thicknesses separating the two jets is $[0.5,4]$.
Global linear stability analysis identifies the most significant  modes driving the changes in the flow dynamics. The neutral lines revealing the critical Reynolds number connected to the presence of the main (steady and unsteady) flow bifurcations, which are presented by global azimuthal modes, show the high complexity of the problem under study, where hysteresis and other types of complex cycles are pointed out.
Finally, the mode interaction is analysed,  highlighting the presence of travelling waves emerging from the interaction of steady states, and the existence of robust heteroclinic cycles that are asymptotically stable. The high level of detail in the results presented, makes this work as a reference for future research development in the field of concentric jets.
\end{abstract}


\section{Introduction\label{sec:introduction}}

Double concentric jets is a configuration enhancing the turbulent mixing of two jets, which is used in several industrial applications where the breakup of the jet into droplets due to flow instabilities is presented as the key technology. Combustion (i.e.: combustion chamber of rocket engines, gas turbine combustion, internal combustion engines, etc.) and noise reduction (e.g.: in turbofan engines) are the two main applications of this geometry, although the annular jets can also be found in some other relevant applications such as ink-jet printers or spray coating.   

\begin{figure}
\begin{center}
\includegraphics[height=0.5\columnwidth]{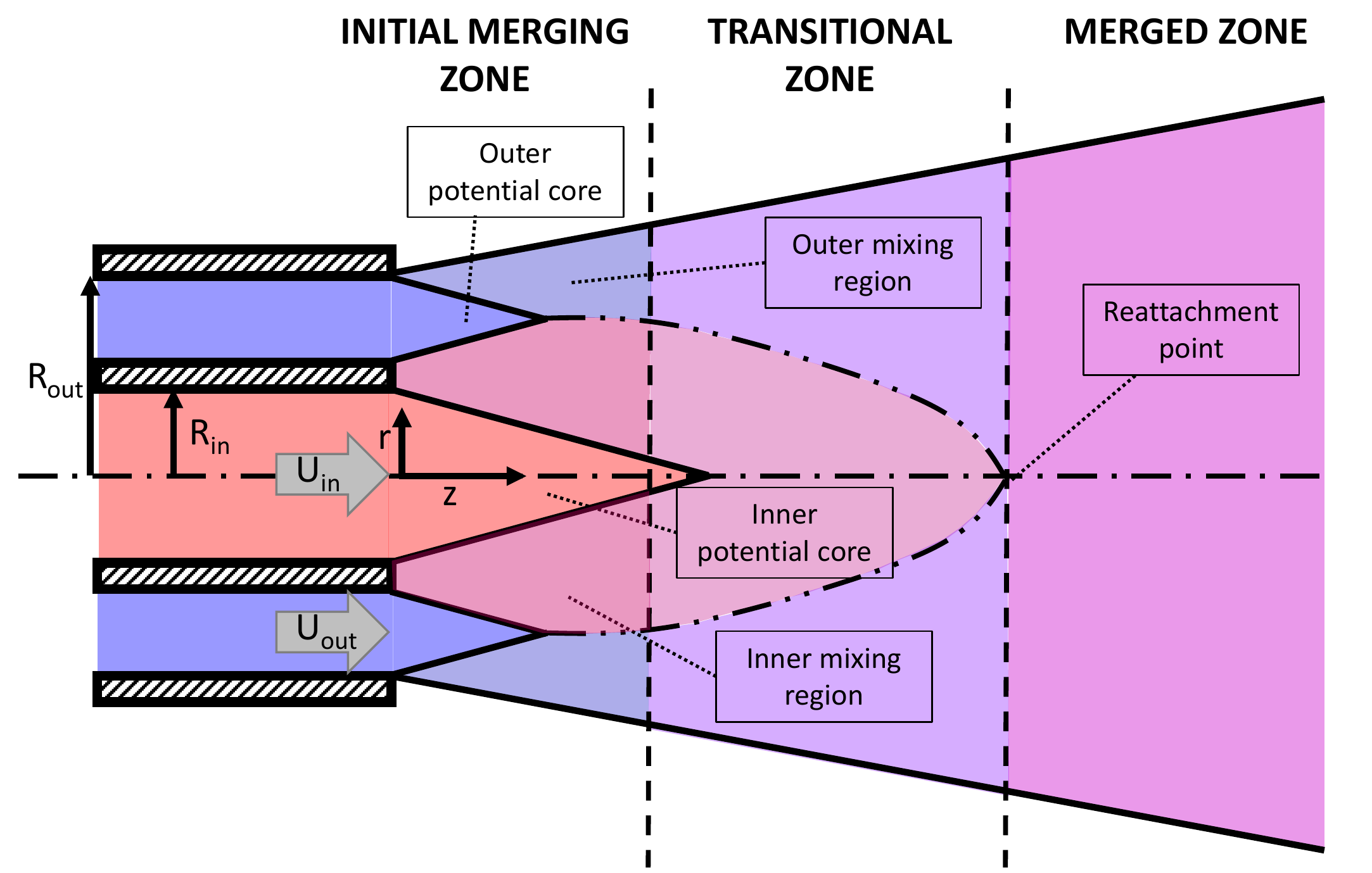}
\end{center}
\caption{ Sketch representing the three flow regimes in the near field of double concentric jets. Figure based on the sketch presented in \cite{KoKwan76,TalamellieGavarini2006}. \label{fig:sketchFlowJet}}
\end{figure}

The qualitative picture emerging from this type of flow divides the inner field of concentric jets in three different regions: (i) initial merging zone, (ii) transitional zone and (iii) merged zone, as presented in fig. \ref{fig:sketchFlowJet}, that follows the initial sketch presented by \cite{KoKwan76}.
In the initial merging zone (i), just at the exit of the two jets, two axisymmetric shear layers (inner and outer boundary layer) develop and start to merge. In this region, we distinguish the inner and outer shear layers, related with the inner and outer jet stream. Then, most of the mixing occurs in the transitional zone (ii), that extends until the external shear layer reaches the centreline. Finally, in the merged zone (iii), the two jets are totally merged, modelling a single jet flow. 
 
Several parameters define the characteristic of this flow: the inner and outer jet velocities, the jet diameters, the shape and thickness of the wall separating both jets, the Reynolds number, the boundary layer state and thickness at the jet exit and the free stream turbulence. Based on these parameters, it is possible to identify several types of flow behaviour, which can be related with the presence of flow instabilities.  

Ko \& Kwan (1976) \cite{KoKwan76} postulated that the double concentric jet configuration could be considered as a combination of single jets. Nevertheless,  Dahm \textit{et al.} (1992) \cite{Dahmetal92} revealed by means of flow visualizations, several topology patterns as function of the outer/inner jet velocity ratio, reflecting that the dynamics of the inner and outer jet shear layers were different from that in a single jet. Moreover, this study exhibited a complex interaction between vortices identified in both shear layers, affecting the instability mechanism of the flow. Buresti \textit{et al.} (1992) \cite{Burestietal92} found that the outer shear layer dominated the flow dynamics for cases in which the outer velocity was much larger than the inner velocity. These authors also detected the presence of an alternate vortex shedding when the wall thickness between the two jets was sufficiently large. The same mechanism was recognised by other authors \cite{Dahmetal92,OlsenKarchmer76}.   Rehab \textit{et al.} (1997) \cite{Rehabetal97} studied in detail the flow differences as function of the outer/inner velocity ratio, finding two different flow regimes when the external jet diameter is much larger than the internal jet one. When the outer/inner velocity ratio was larger than a critical value, the authors spotted a low frequency recirculation bubble at the jet outlet. On the contrary, for outer/inner velocity ratio smaller than such critical value, the outer (still fast) jet excites the inner jet, which ends oscillating at the same frequency as the external jet. This is known as the lock-in phenomenon. Moreover, the oscillation frequency detected was similar to the one defined by a Kelvin-Helmholtz flow instability, which is generally encountered in single jets. This lock-in phenomenon was also identified by other authors \cite{Dahmetal92,DaSilvaetal2003}.

Following previous works \cite{Burestietal92,Dahmetal92,OlsenKarchmer76} and paying especial attention to the separating wall thickness and the vortex shedding located behind the wall,  Wallace \& Redekopp (1992) \cite{WallaceRedekopp92} showed that the wall thickness and sharpness change the characteristic of the jet. Segalini \& Talamelli (2011) \cite{SegaliniTalamelli2011} performed experiments to inspect in detail the effects of the outer/inner velocity ratio and the wall thickness in double concentric jets. These authors found that for small outer/inner velocity ratios, the inner jet presents its own flow instability in the shear layer, while a different flow instability was identified in the outer jet. On the contrary, for large outer/inner velocity ratios, the outer shear layer drives the flow dynamics, forcing the inner shear layer to oscillate with the same frequency, occurring in the lock-in phenomenon previously mentioned. Finally, for similar outer/inner velocity ratios, a Von Kármán vortex street was detected near the separating wall, as also depicted by other authors \cite{Burestietal92,Dahmetal92,OlsenKarchmer76}. A wake instability affected the inner and outer shear layers, reversing the lock-in phenomenon.

Different configurations can also be found, changing the velocity ratio between jets.  Williams \textit{et al.} (1969) \cite{williams69} worked on the influence of the exterior/interior velocity ratio on noise attenuation, which was analysed experimentally. It was observed that for some given configurations, more noise attenuation was present than for the others, with a maximum between $12$ and $15 dB$. 

Talamelli \& Gavarini (2006) \cite{TalamellieGavarini2006} performed a local linear stability analysis, finding that for specific wall thickness, the vortex shedding identified behind the wall, can be related with an absolute instability that exists for some specific outer/inner velocity ratios. The authors explained that this absolute instability may trigger the destabilization of the flow field. This theoretical work was verified experimentally by  \cite{Orluetal2008}. These authors showed once more that the wake behind the wall separating the two jets creates a vortex shedding driving the frequency of the external shear layer also controlling the evolution of the inner shear layer, which can be the mechanism that triggers a global absolute instability. This passive mechanism can be considered as a potential tool for flow control, delaying the transition to turbulence by means of controlling the near field of the jet. Recently,  Canton \textit{et al.} (2017) \cite{Cantonetal2017} performed a global linear stability analysis to study more in detail this vortex shedding mechanism behind the wall. They examined a concentric jet configuration with a very small wall thickness ($0.1D_i$, with $D_i$ the inner jet diameter), but the authors selected an outer/inner velocity ratios where it was known that the alternate vortex shedding behind the wall was driving the flow. A global unstable mode (absolute instability) with azimuthal wavenumber $m=0$ was found, confirming that the primary instability was axisymmetric (the modes with $m=1,2$ were stable at the flow conditions at which the study was carried out). The highest intensity of the global mode was located in the wake of the jet, composed by an array of counter-rotating vortex rings. The shape of the mode  changes when moving along its neutral curve, revealing through the numerical simulations a Kelvin-Helmholtz instability over the shear-layer between the two jets and in the outer jet at high Reynolds numbers. Nevertheless, the authors showed that the wavemaker was located in the bubble formed upstream the separating wall, in good agreement with the results presented by \cite{Tammisola2012}, who performed a similar stability analysis in a two-dimensional configuration (wakes with co-flow). 

The stability of annular jets, a limit case where the inner jets have zero velocity, has also been investigated. In different analysis of annular jets \cite{bogulawski,michalke1999}, it has been illustrated that this type of axisymmetric configuration does not behave as it appears. The $m=0$ modes studied have been shown to be stable, and the dominant mode found by both studies is helical ($m=1$). In addition, to characterise the annular jet, these investigations analyse the behaviour of the case by adding an azimuthal component to the inflow velocity, making the discharge of the annular jet eddy-like, comparing the evolution of the frequency and growth rate of this $m=1$ mode.

This paper expands on the work done by \cite{Cantonetal2017}, where they use a specific geometry and vary the outer/inner velocity ratio. This paper presents a complete characterisation of the main global modes identified in two concentric jets. The wall thicknesses separating the two jets are defined in the interval $L\in [0.5,4]$, and the flow is simulated for different inside/outside velocity ratios in the interval $U_i/U_o\in[0,2]$, where the case with $U_i/U_o=0$ represents an annular jet. Global modes with azimuthal wavenumber $m=0$ (axisymmetric modes), $m = 1$ and $m = 2$ will be searched for. As identified in the literature \cite{michalke1999,bogulawski}, no axisymmetric modes ($m=0$) could be identified for any of the distances, as this is a helical case. This paper expands the conclusions found in these two previous works, extending the results to different wall thicknesses between jets. This part of the paper studies in detail the configuration of two concentric jets at low Reynolds numbers. Using a linear approximation of the equations that model the flow, the base flows will be obtained on which to apply the linear stability analysis, by means of which it is possible to identify the most relevant modes that influence the flow dynamics.

This work also performs a study of mode selection, as some configurations presents interactions between different modes. Different analysis have been done to know the different coherent structures when there is an interaction between modes. \cite{sierra2022Lorite} conducted the study on the flow past a rotating sphere, finding different coherent structures on a triple-Hopf bifurcation. Some of these configurations are steady states, travelling waves or rotating waves. 

To the authors' knowledge, this is the first time that the characterisation of two concentric jets is presented with such level of detail, presenting neutral curves for a wide range of different configurations, as well as providing a deep understanding of the flow physics through the interaction between the different modes. 

The article is organized as follows. Section \ref{sec:geometryNS} defines the problem and the governing equations for the double concentric jets, as well as the linear stability equations and the methodology for mode selection. The axisymmetric steaty-state is characterised in Section \ref{sec:SteadyState}. In Section \ref{sec:LST}, we perform a parametric exploration in terms of the velocity ratio between the jets and the jet distance in order to determine the neutral curves of global stability. The results about the mode selection are discussed in Section  \ref{sec:ModeInt}. Finally, Section \ref{sec:conclusions} summarises the main conclusions.

\section{Problem formulation\label{sec:geometryNS}}

\subsection{Computational domain and general equations}

\begin{figure}
    \centering
    \includegraphics[height=0.5\columnwidth]{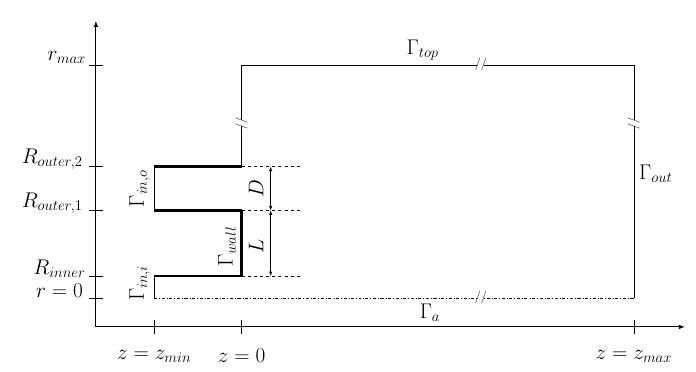}
    \caption{Computational domain of the configuration of two concentric jets, used in StabFem.}
    \label{fig:comp_domain}
\end{figure}

The computational domain, represented in \cref{fig:comp_domain}, models a coaxial flow configuration, which is composed of two inlet regions, an inner and outer pipe, both of diameter $D$ and length $5D$, i.e. $z_{min} = -5 D$. The computational domain has an extension of $z_{max} = 50 D$ and $r_{max} = 25 D$. The distance between the pipes is equal to $L$, measured from the inner face of the outer tube to the face of the inner jet.

The governing equations of the flow within the domain are the incompressible Navier--Stokes equations. These are written in cylindrical coordinates $(r,\theta,z)$, which are made dimensionless by considering $D$ as the reference length scale and $W_{o,max}$ as the reference velocity scale, which is the maximum velocity in the outer pipe at $z=z_{min}$. 

\begin{subequations}
\begin{align}
& \frac{\partial \vec{U}}{\partial t} + \vec{U}\cdot \nabla\vec{U} = -\nabla P +  \nabla \cdot \tau(\vec U ),  \qquad \nabla \cdot \vec U = 0, \\
&\text{with }  \tau(\vec{U}) = \frac{1}{\Rey}  (\nabla \vec U + \nabla \vec U^T), \qquad \Rey = \frac{W_{o,max} D}{\nu}.
\label{eq:ScalesINS}
\end{align}
\label{eq:INS}
\end{subequations}
The dimensionless velocity vector $\vec{U} = (U,V,W)$ is composed of the radial, azimuthal and axial components, $P$ is the dimensionless-reduced pressure, the dynamic viscosity $\nu$ and the viscous stress tensor $ \tau(\vec U )$.

The incompressible Navier--Stokes equations \cref{eq:INS} are complemented with the following boundary conditions  
\be{BoundaryConditions}
\vec{U} = (0, 0, W_{i})  \text{ on } \Gamma_{in,i} \text{ and } \vec{U} = (0, 0, W_{o}) \text{ on }  \Gamma_{in,o},
\ee
where $$W_{i}=\delta_u \tanh{ \big(b_i (1 - 2r)\big)} \text{ and }  W_{o}=\tanh \left[  b_o \left( 1 + \left\lvert \frac{r- (R_{outer,1}+R_{outer,2})}{D} \right\rvert \right)  \right].$$
The parameter $\delta_u$ corresponds to the velocity ratio between the two jets, defined as $\delta_u=W_{i,max}/W_{o,max}$. The parameters $b_o$ and $b_i$ represent the boundary layer thickness within the nozzle, which are fixed equal to $5$ (as in  \cite{Cantonetal2017}). There is a weak influence of the boundary layer thickness on the stability properties of the jet, and it is related to the vortex shedding regime developed upstream the separation wall (more details may be found in  \cite{TalamellieGavarini2006}). Finally, no-slip boundary condition is set on $\Gamma_{wall}$ and stress-free ($\big(\frac{1}{Re} \tau(\vec{U})-P\big) \cdot \vec{n} ={\bf 0}
$) boundary condition is set on $\Gamma_{top}$ and $\Gamma_{out}$, as shown in \cref{fig:comp_domain}. 

In the sequel, Navier--Stokes equations \cref{eq:INS} and the associated boundary conditions will be written symbolically under the form 
\be{GoverningEquationsNSCompact}
  \vec{B}  \frac{\partial \vec{Q}}{\partial t} = \vec{F}(\vec{Q}, \vec{\bm}) \equiv \vec{L} \vec{Q}  + \vec{N}(\vec{Q},\vec{Q}) + \vec{G} (\vec{Q},\bm{\eta}),
\ee
with the flow state vector $\vec{Q} = \left[ \vec{U},{P} \right]^T$, $\bm{\eta} = \left[\Rey, \delta_u \right]^T$. 
Such a form of the governing equations takes into account a linear dependency on the state variable $\vec{Q}$ through $\vec{L}$. And a quadratic dependency on the parameters and the state variable through operators $\vec{G}(\cdot, \cdot)$ and $\vec{N}(\cdot, \cdot)$.

\subsection{Asymptotic stability}

\subsubsection{Linear stability analysis}
In this study, the authors attempt to characterize the stable asymptotic state from the spectral properties of the Navier--Stokes equations \cref{eq:INS}. First, let us consider the stability of an axisymmetric steady-state solution named $\QVBF$, which will be also referred to as \textit{trivial steady-state}. For that purpose, let evaluate a solution of \cref{eq:INS} in the neighbourhood of the trivial steady state, i.e., a perturbed state as follows,
\be{AnsatzLinearStab}
\vec{Q}(\vec{x}, t) = \QVBF(\vec{x}, t) + \varepsilon \hat{\vec{q}}(r, z) \text{e}^{-\im (\omega t - m \theta)}.
\ee
The next step consists in the characterization of the dynamics of small-amplitude perturbations around this base flow by expanding them over the basis of linear eigenmodes (\ref{eq:AnsatzLinearStab}).
If there is a pair $\left[\im \omega_\ell, \hat{\vec{q}}_{\ell} \right]$ with $\text{Im} (\omega_\ell) > 0$ (resp. the spectrum is contained in the half of the complex plane with negative real part)  there exists a basin of attraction in the phase space where the trivial steady-state $\QVBF$ is unstable (resp. stable) \cite{kapitula2013spectral}. The eigenpair $\left[\im \omega_\ell, \hat{\vec{q}}_{\ell} \right]$ is determined as a solution of the following  eigenvalue problem,
\be{FirstOrderB}
\displaystyle \vec{J}_{(\omega_\ell, m_\ell)} \qMVZ{\ell} = \Big(i\omega_\ell \vec{B} - \frac{\partial \Fop}{\partial \vec{q} }|_{\vec{q} = \QVBF, \parV = \bm{0}} \Big) \qMVZ{\ell},
\ee
where $\Big(\frac{\partial \Fop }{\partial \vec{q} }|_{\vec{q} = \QVBF, \parV = \bm{0}} \Big) \qMVZ{\ell} = \vec{L}_{m_\ell} \qMVZ{\ell} + \vec{N}_{m_\ell}(\QVBF, \qMVZ{\ell} ) + \vec{N}_{m_\ell}(\qMVZ{\ell}, \QVBF )$. The subscript $m_\ell$ indicates the azimuthal wavenumber used for the evaluation of the operator. 
In the following, we account for eigenmodes $\qMVZ{\ell}(r,z)$ that have been normalised in such a way $\langle \vec{\hat u}_{(z_\ell)}, \vec{\hat u}_{(z_\ell)} \rangle_{L^{2}} = 1$.

\subsubsection{Methodology for the study of mode selection}
\label{sec:ModeSelection}
In the following, we briefly outline the main aspects of the methodology employed in the study of mode interaction, a comprehensive explanation is left to \cref{sec:Appendix_reductionprocedureCoefficients}.
The determination of the attractor or coherent structure is explored within the framework of equivariant bifurcation theory. The trivial steady-state is axisymmetric, i.e. the symmetry group is the orthogonal group $O(2)$. Near the onset of the instability, dynamics can be reduced to those of the centre manifold. Particularly, due to the non-uniqueness of the manifold one can always look for its simplest polynomial expression, which is known as the \textit{normal form} of the bifurcation. The reduction to the normal form is carried out via a multiple scales expansion of the solution $\QV$ of \cref{eq:GoverningEquationsNSCompact}. The expansion considers a two scale development of the original time $t \mapsto t + \varepsilon^2 \tau$, here $\varepsilon$ is the order of magnitude of the flow disturbances, assumed to be small $\varepsilon \ll 1$. In this study we carry out a normal form reduction via a weakly non-linear expansion, where the small parameters are
$$\varepsilon_{\delta_u}^2 =  \delta_{u,c} - \delta_{u} \ \sim \varepsilon^2  
\text{ and } \varepsilon_{\nu}^2 = \big( \nu_c - \nu \big) = \big( \Rey_c^{-1} - \Rey^{-1} \big) \sim \varepsilon^2.$$
A fast timescale $t$ of the self-sustained instability and a slow timescale of the evolution of the amplitudes $z_i(\tau)$ are also considered in \cref{eq:dynsyst}, for $i=1,2,3$.
The ansatz of the expansion is as follows
\be{Normal_Form_Ansatz_WNL2}
\displaystyle \vec{Q}(t,\tau)  = \QVBF  + \varepsilon \vec{q}_{(\varepsilon)}(t,\tau) + \varepsilon^2  \vec{q}_{(\varepsilon^2)}(t,\tau)  + O(\varepsilon^3).
\ee
Herein, we evaluate the mode interaction between two steady symmetry breaking states with azimuthal wave number $m_1 = 1$ and $m_2 = 2$, that is,
\bes{AnsatzVel}
    \vec{q}_{(\varepsilon)}(t,\tau) & = \big(  z_1(\tau) \qMVZ{1}(r,z) e^{-\im m_1 \theta} + \text{c.c.} \big ) \\
   & +  \big(   z_2(\tau) \qMVZ{2}(r,z) e^{-\im m_2 \theta} + \text{c.c.} \big).
\ees
Note that the expansion of the LHS of \cref{eq:GoverningEquationsNSCompact} up to third order is as follows 
\be{GoverningEqExpandedLHS}
\displaystyle  \varepsilon \vec{B} \frac{\partial \vec{q}_{(\varepsilon)}} {\partial t} + \varepsilon^2 \vec{B} \frac{\partial \vec{q}_{(\varepsilon^2)}} {\partial t} 
+ \varepsilon^3 \big[ \vec{B} \frac{\partial \vec{q}_{(\varepsilon^3)}} {\partial t}\big]  +  O(\varepsilon^4),
\ee
and the RHS respectively,
\be{GoverningEqExpandedRHS}
\vec{F} (\vec{q}, \bm{\eta}) = \vec{F}_{(0)} + \varepsilon \vec{F}_{(\varepsilon)}  + \varepsilon^2 \vec{F}_{(\varepsilon^2)}  + \varepsilon^3 \vec{F}_{(\varepsilon^3)} + O(\varepsilon^4).   
\ee
Then, the problem up to third order in $z_1$ and $z_2$ can be reduced to \cite{armbruster1988heteroclinic}
\bes{dynsyst}
    \dot{z}_1 &= \lambda_1 z_1 + e_3 \overline{z}_1 z_2 + z_1\big( c_{(1,1)} |z_1|^2 +  c_{(1,2)} |z_2|^2 \big), \\
     \dot{z}_2 &= \lambda_2 z_2 + e_4 z^2_1 + z_2\big( c_{(2,1)} |z_1|^2 +  c_{(2,2)} |z_2|^2 \big). \\
\ees
An exhaustive analysis of the nonlinear implications of this normal form on dynamics is left to \cref{sec:ModeInt}. The procedure followed for the determination of the coefficients $c_{(i,j)}$ for $i,j=1,2$ and $e_3$ and $e_4$ is left to \Cref{sec:Appendix_reductionprocedureCoefficients}.

\subsubsection{Numerical methodology for stability tools}

Results presented herein follow the same numerical approach adopted by \cite{fabre2018practical, sierra2020efficient, sierra2020bifurcation, sierra2022Lorite}, where a comparison with DNS can be found. The calculation of the steady-state, the eigenvalue problem and the normal form expansion are implemented in the open-source software \texttt{FreeFem++}. Parametric studies and generation of figures are collected by \texttt{StabFem} drivers, an open-source project available in \url{https://gitlab.com/stabfem/StabFem}. For steady-state, stability and normal form computations, we set the \textit{stress-free} boundary condition at the outlet, which is the natural boundary condition in the variational formulation.  

The resolution of the steady nonlinear Navier-Stokes equations is tackled by means of the Newton method. While, the generalised eigenvalue problem (\cref{eq:FirstOrderB}) is solved following the Arnoldi method with spectral transformations. The normal form reduction procedure of \cref{sec:ModeSelection} only requires to solve a set of linear systems, which is also carried out within \texttt{StabFem}. On a standard laptop, every computation considered below can be attained within a few hours.

\section{Characterisation of the axisymmetric steady-state}
{\label{sec:SteadyState}
\subsection{Velocity ratio effects}
We begin by characterizing the development of the axisymmetric steady-state with varying $\delta_u$ at a constant Reynolds number fixed to $\text{Re}=100$. \Cref{fig:BaseFlowContours} synthesises the main topological changes experiences by the steady-state. At $\delta_u = 0$, the solution (point (a) in \cref{fig:BaseFlowContours}) represents an annular jet, which diffuses as it travels downstream and enters the ambient fluid. This figure illustrates that the solution curve can be divided into three segments. The first segment comprised between $0 \leq \delta_u < \delta_u^1$ is characterised by an inner jet nearly trapped by a large recirculation region with a characteristic length $L_r$, which remains almost constant with the velocity ratio. 

\begin{figure}
    \centering
    {\includegraphics[width=0.9\columnwidth]{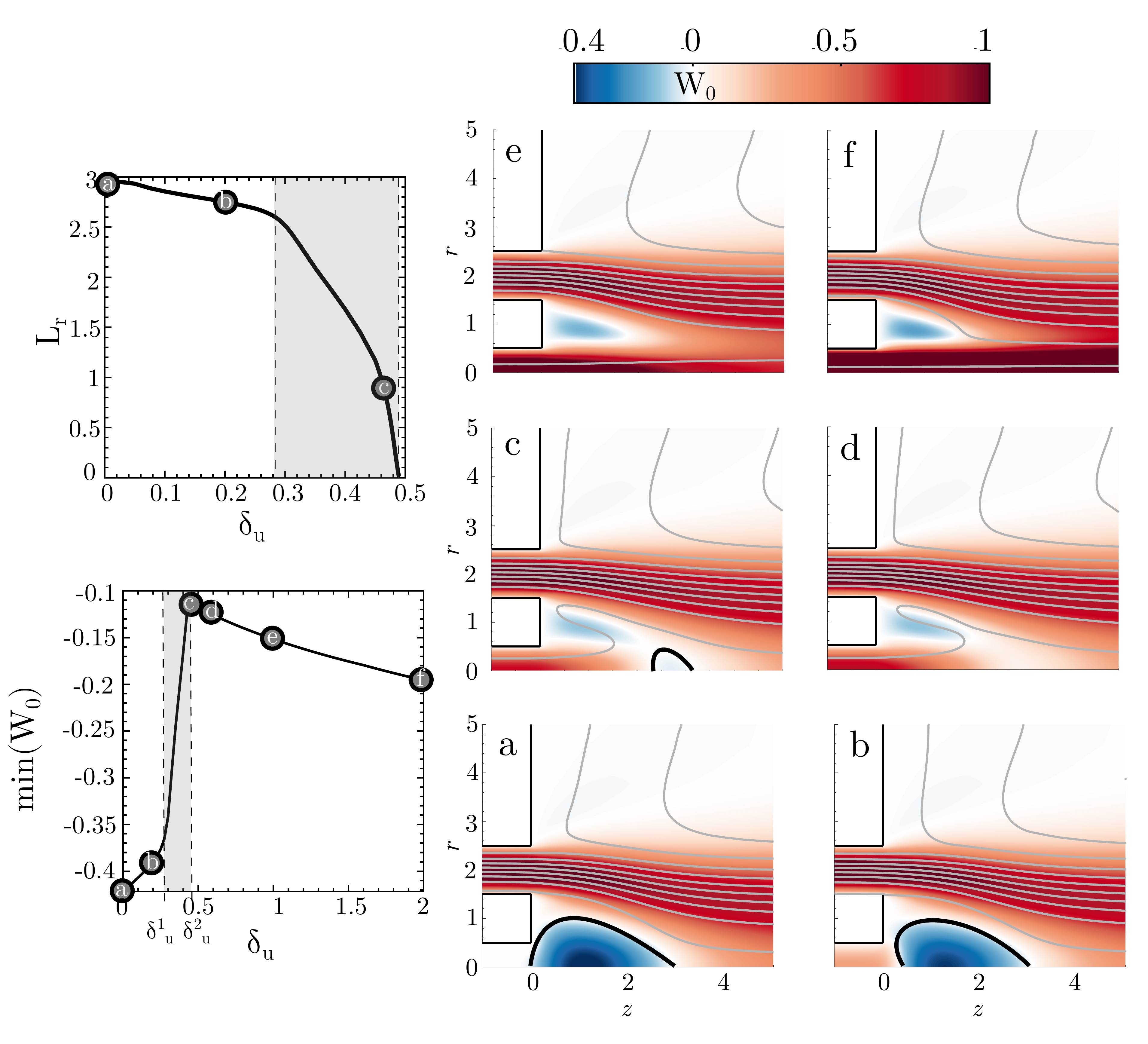}} \quad
    \caption{ Evolution of the recirculation length ($L_r$) of the recirculating bubble with respect to the velocity ratio $\delta_u$ between the inner and outer jet. The diagram of the second row on the left displays the minimum value within the domain of the axial velocity. It is spatially localised within the recirculating region for $\delta_u < 0.5$ and near the middle wall for larger values of the velocity ratio. Meridional projections of the axisymmetric streamfunction isolines and the axial velocity contour in a range of $(z,r) \in [-1,5] \times [0,5]$. }
    \label{fig:BaseFlowContours}
\end{figure}

In the second region, which ranges between $\delta_u^1 < \delta_u < \delta_u^2$ and it is represented as a shaded area in the figure, the recirculating region rapidly reduces its size. In this region, the axial velocity of the inner jet is comparable with the axial velocity observed in the recirculating region, which promotes mixing between both regions. As the velocity ratio is increased, the inner jet is sufficiently energetic to break the recirculating region, which occurs between point (c) and (d) in \cref{fig:BaseFlowContours}. The final segment, that ranges between $\delta_u > \delta_u^2$, is characterised by two quasi-planar jets that rapidly mix to form a larger one at around $z \approx 5$. 

\section{Linear stability analysis \label{sec:LST}}

 We explore the parameter space ($\text{Re}$, $\delta_u$, $L$). Herein, we examine the velocity ratio between the jets ($ 0 < \delta_u < 2$) and the distance between the jets ($0.5 < L < 4$). Within this range of parameters, we have analysed the linear stability properties of the flow configuration.
For this purpose, we first investigate the influence of the jet distance on the stability for the case of the annular jet ($\delta_u=0$). 

\begin{figure}
    \centering
    \subfloat[]{\includegraphics[width=0.235\columnwidth]{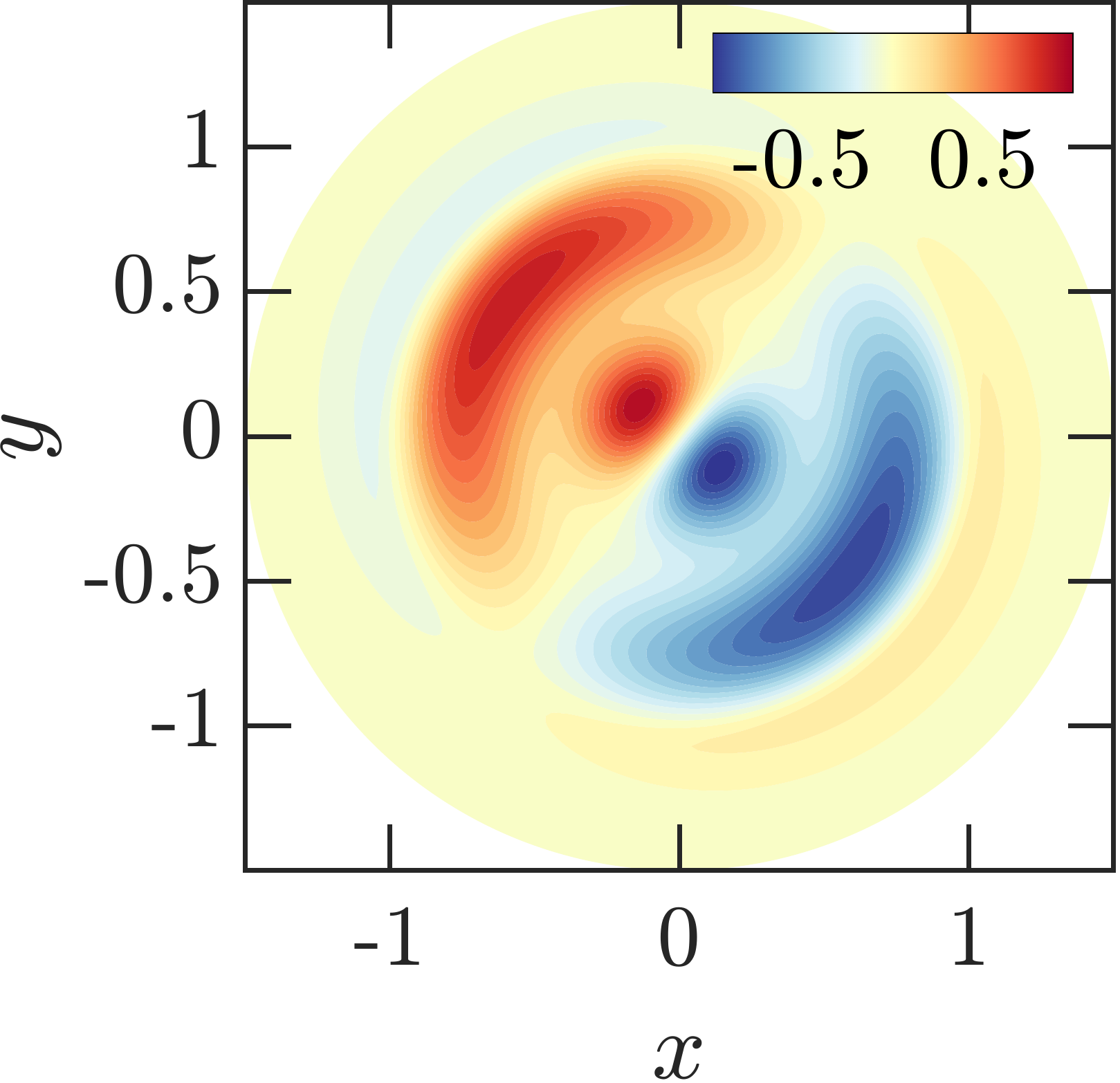}} \quad
        \subfloat[]{\includegraphics[width=0.235\columnwidth]{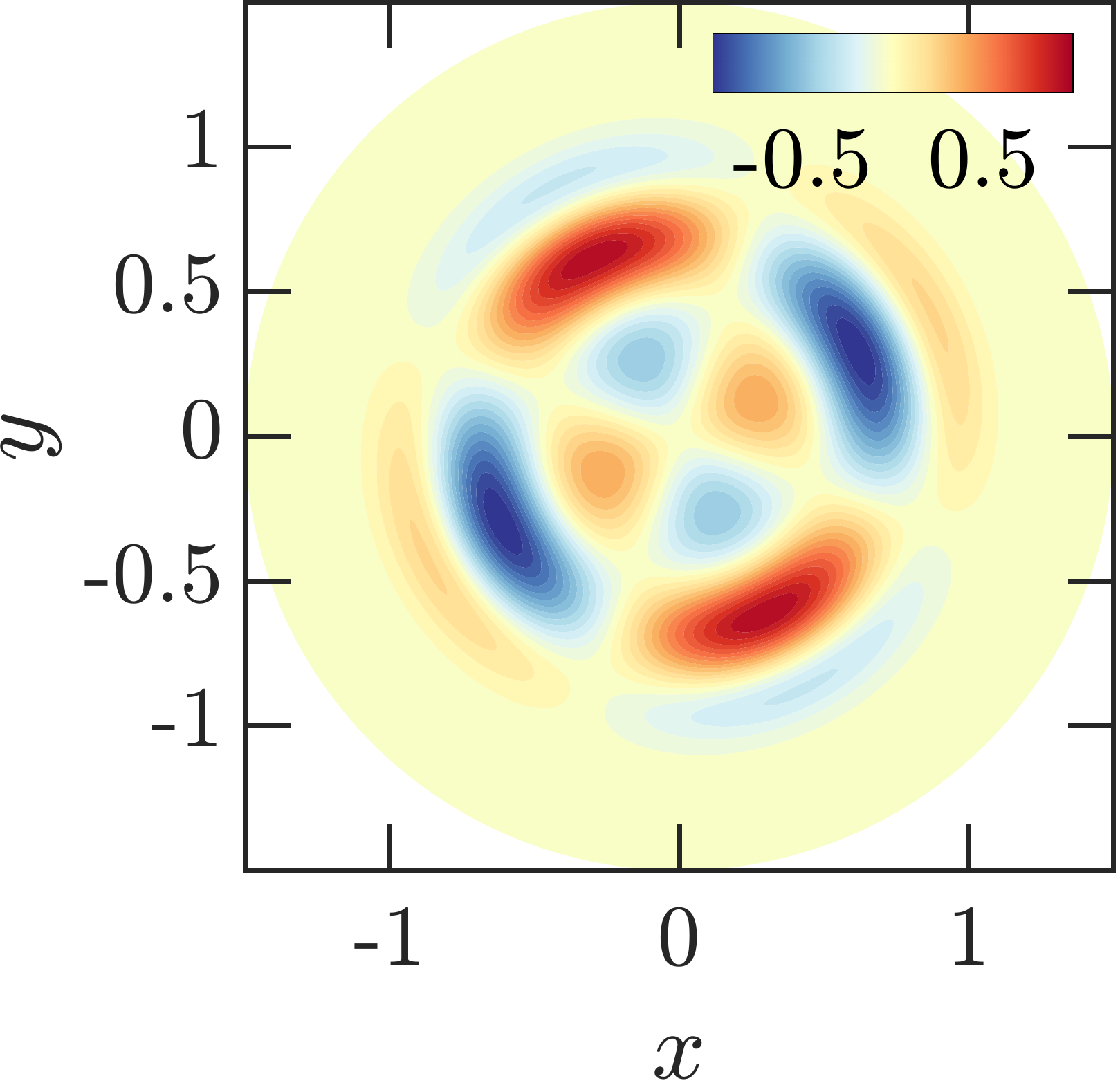}} \quad
        \subfloat[]{\includegraphics[width=0.22\columnwidth]{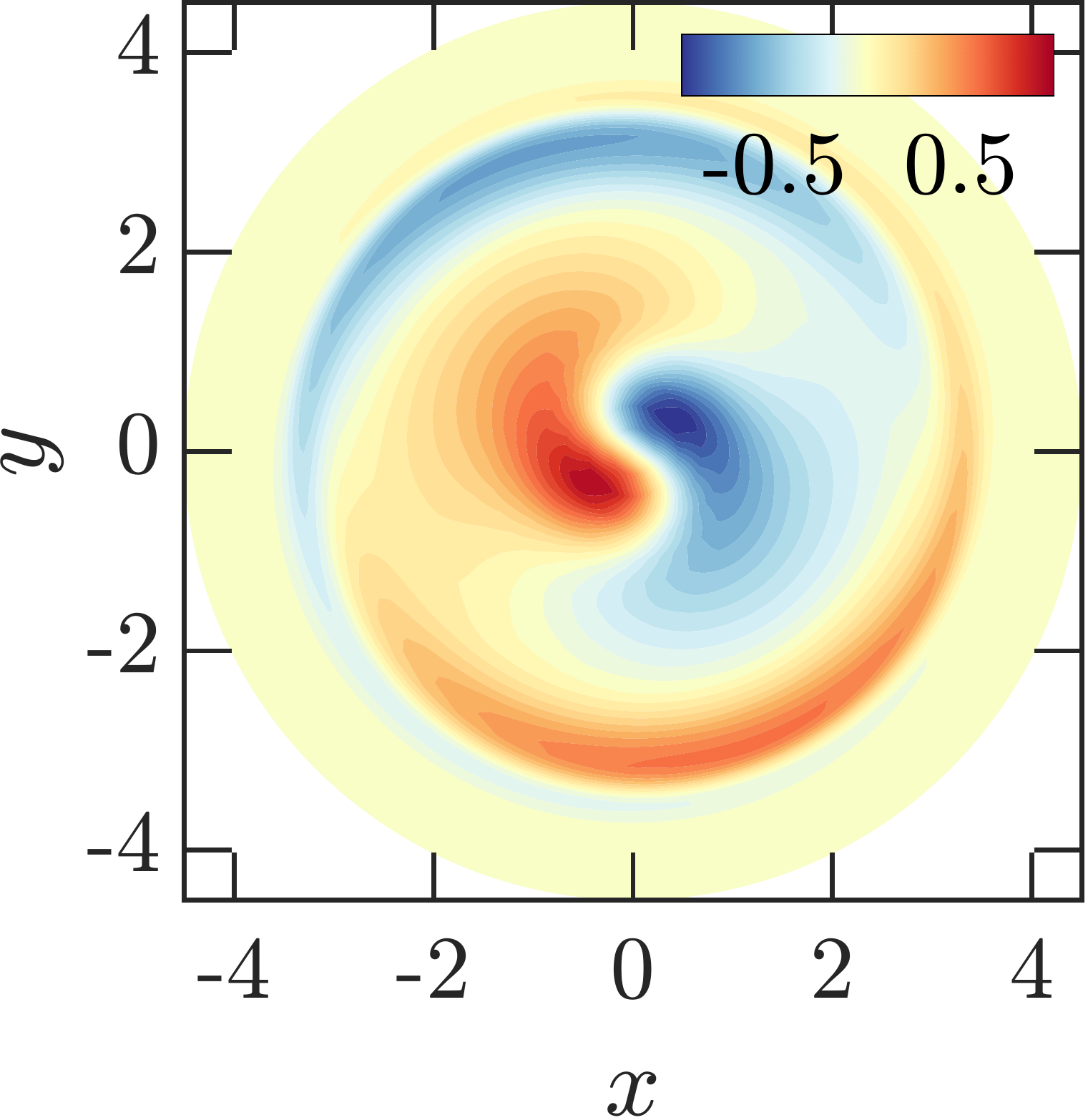}} \quad
        \subfloat[]{\includegraphics[width=0.22\columnwidth]{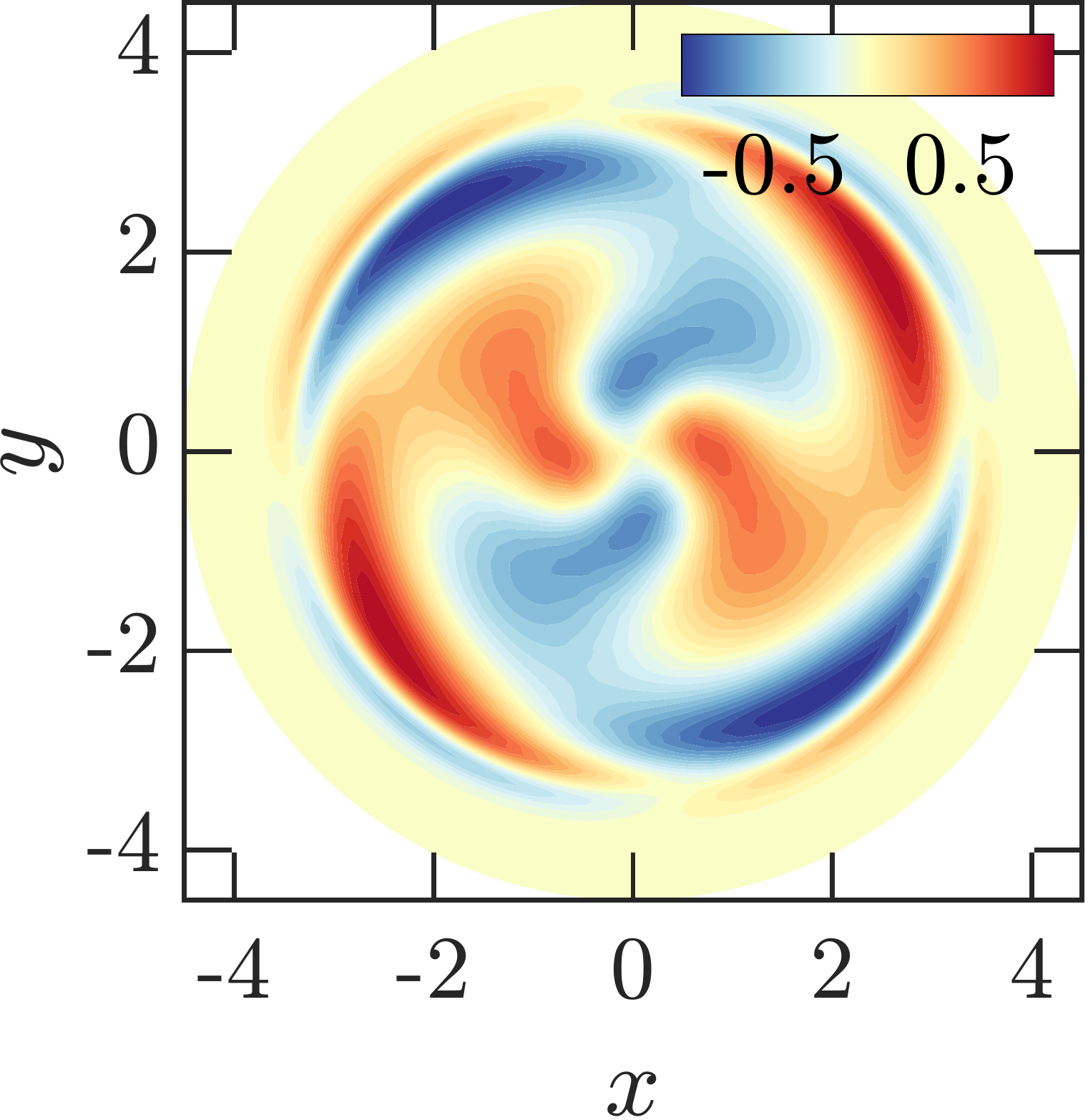}}
        \\
    \subfloat[]{\includegraphics[width=0.45\columnwidth]{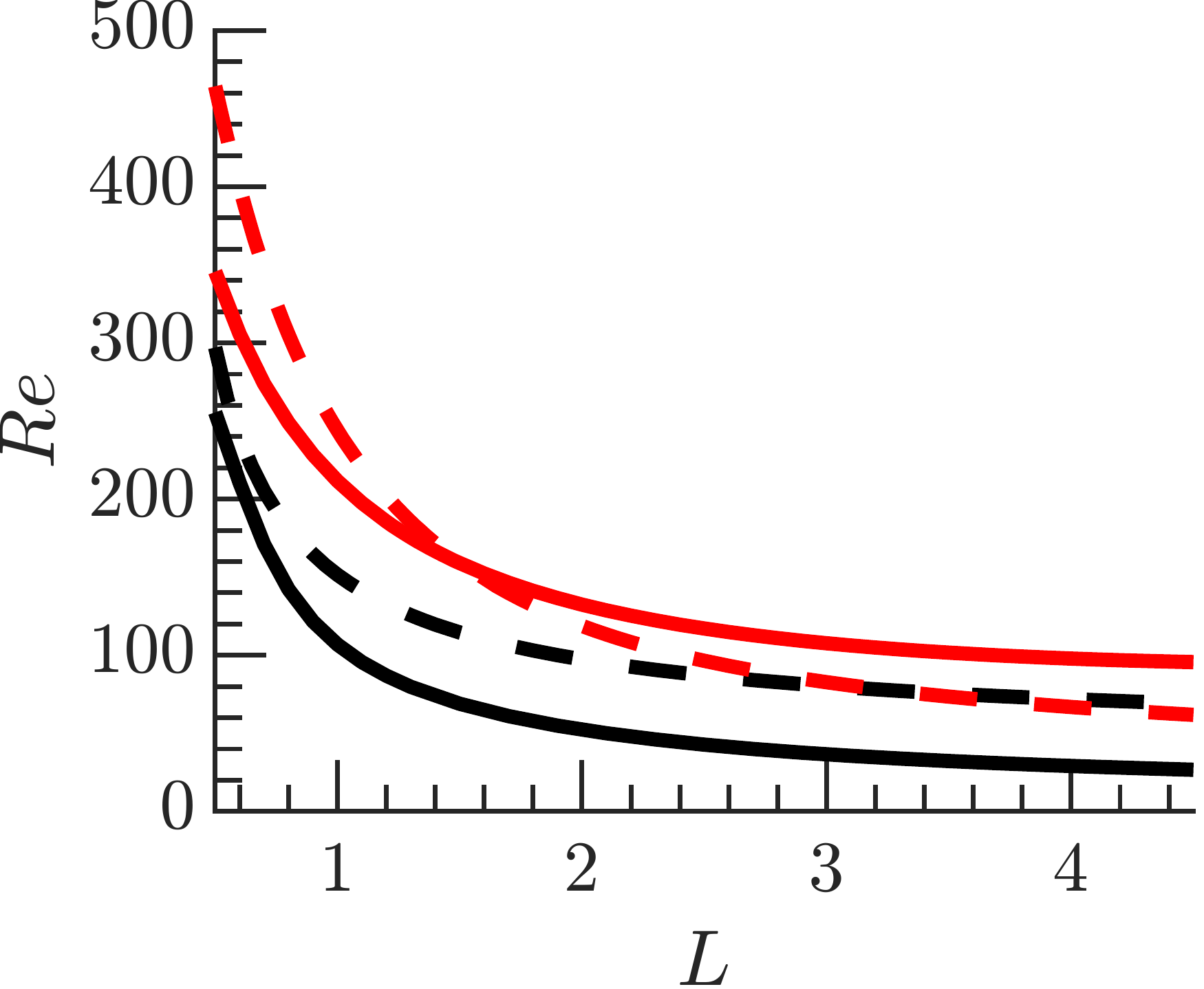} }
    \subfloat[]{\includegraphics[width=0.45\columnwidth]{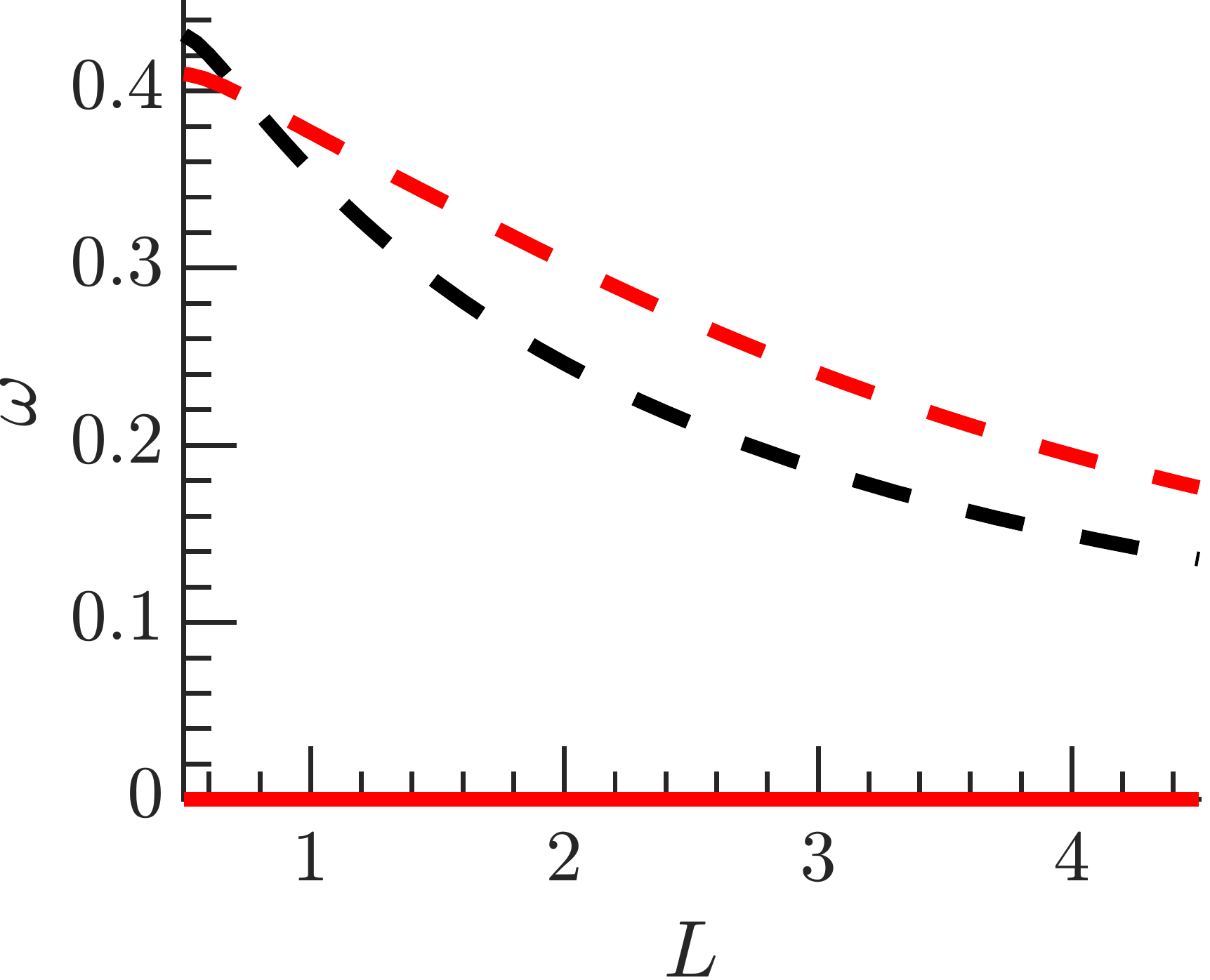} }
    \\
    \subfloat[]{\includegraphics[width=0.45\columnwidth]{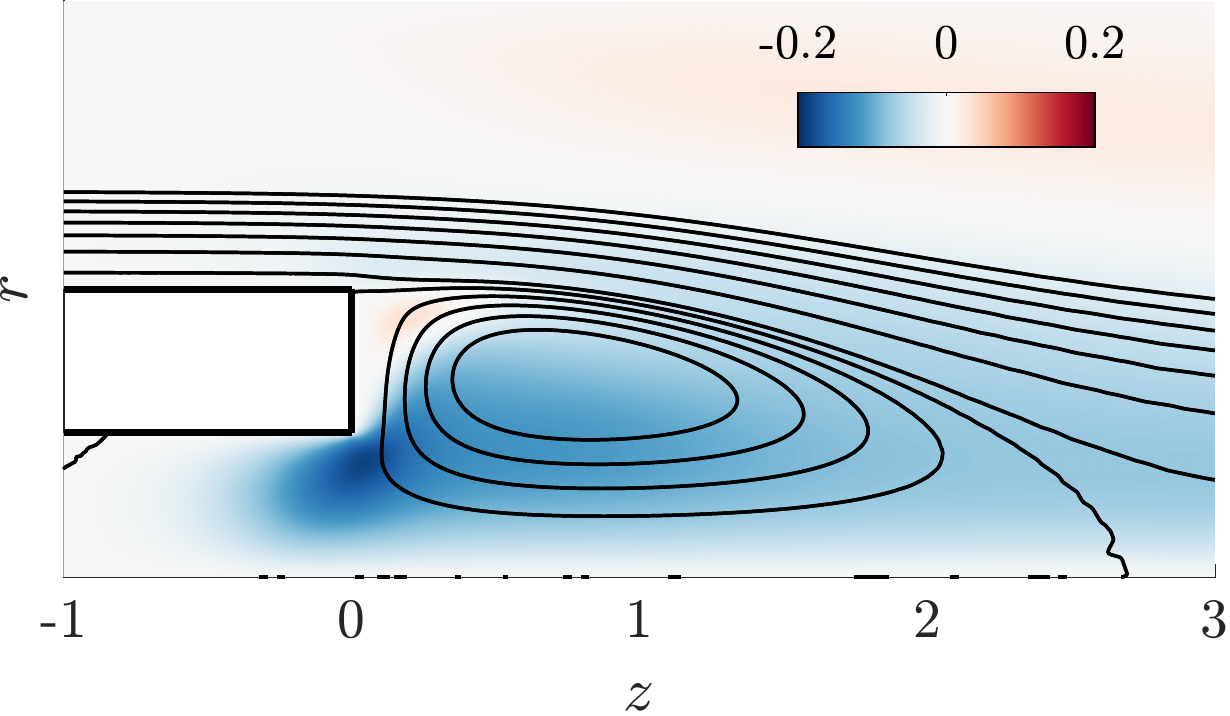}} \quad
    \subfloat[]{\includegraphics[width=0.45\columnwidth]{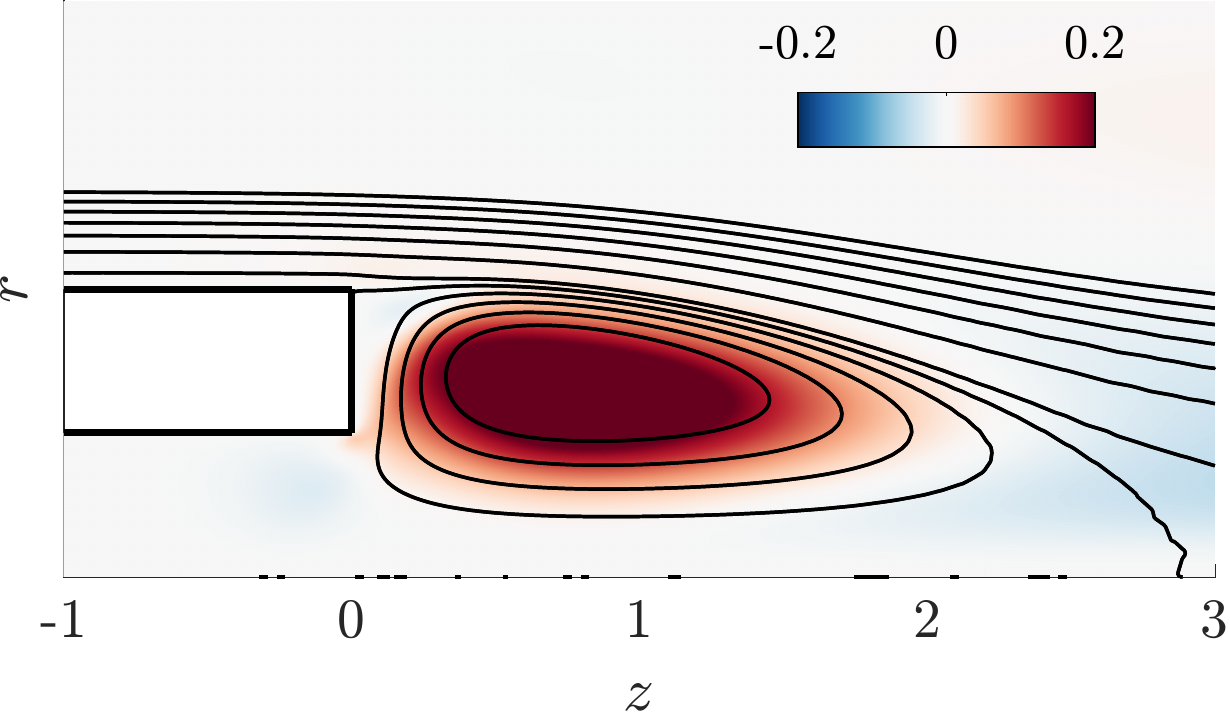}}
    \\
    \subfloat[]{\includegraphics[width=0.75\columnwidth]{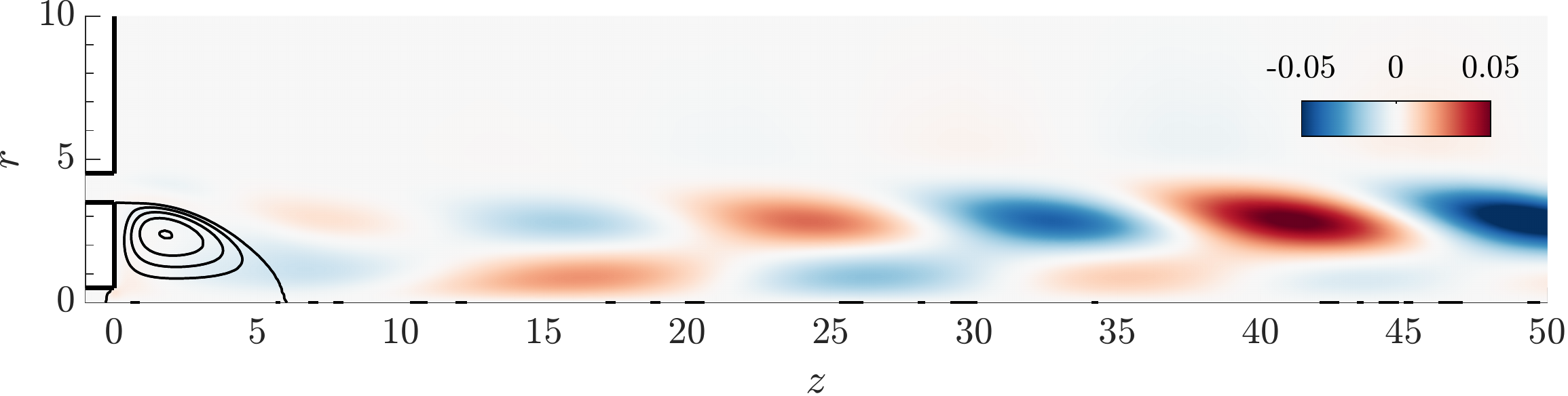}} \\
    \subfloat[]{\includegraphics[width=0.75\columnwidth]{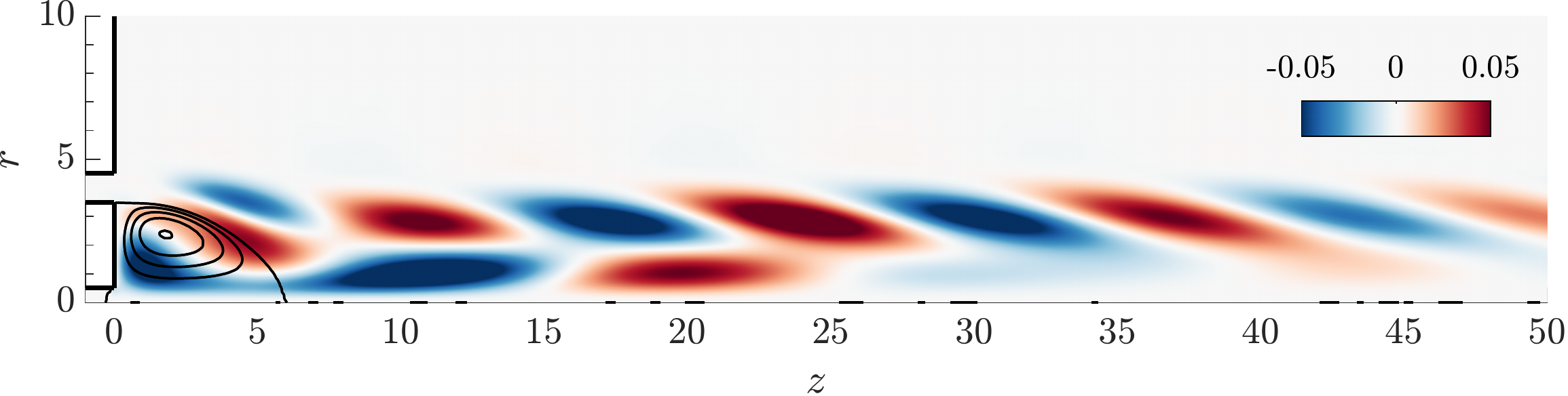}}
    \caption{ Cross-section view at $z=1$ of the four unstable modes at criticality for the annular jet case ($\delta_u=0$). The streawise component of the vorticity vector $\varpi_z$ is visualised by colours. (a) Mode $S_1$ for $L=0.5$, (b) Mode $S_2$ for $L=0.5$, (c) Mode $F_1$ for $L=3$ and (d) Mode $F_2$ for $L=3$.
    (e) Linear stability boundaries for the annular jet ($\delta_u = 0$). (f) Frequency evolution of the unsteady modes. Legend: $S_1$ mode is displayed with a solid black line, $S_2$ with a solid red line and $F_1$ and $F_2$ modes are depicted with dashed black and red lines, respectively.
    Streamwise velocity of the neutral modes for $L=3$ and $\delta_u=0$ (i) $F_1$, (h) $F_2$ .}
    \label{fig:NeutralCurve_NoCoaxialFlow}
\end{figure}

These findings are summarized in \cref{fig:NeutralCurve_NoCoaxialFlow} which displays the evolution of the critical Reynolds number with respect to the distance ($L$) for the four most unstable modes: two steady modes with azimuthal wavenumber $m=1$ and $m=2$, hereinafter referred to as modes $S_1$ and $S_2$, respectively. A cross-section view at $z=1$ is displayed in \cref{fig:NeutralCurve_NoCoaxialFlow} (a-b). The other two unsteady modes, named $F_1$ and $F_2$ have respectively azimuthal wavenumbers $m=1$ and $m=2$. A cross-section view of these two modes is displayed in \cref{fig:NeutralCurve_NoCoaxialFlow} (c-d). Please note that for the chosen set of parameters the axisymmetric unsteady mode $F_0$, is always found at larger Reynolds numbers than the aforementioned modes. This is one of the major differences with the case studied by \cite{Cantonetal2017}, for small values of the jet distance $L$, the dominant instability is an unsteady axisymmetric one, which would be named $F_0$ with our nomenclature. Thus, in the following, we only include the results for the $S_1$, $S_2$, $F_1$ and $F_2$ modes. The primary instability of the annular jet is then a steady symmetry-breaking bifurcation that leads to a jet flow with a single symmetry plane, displayed in \cref{fig:NeutralCurve_NoCoaxialFlow} (a). On the contrary, bifurcations that lead to the mode $S_2$ possess two orthogonal symmetry planes, see \cref{fig:NeutralCurve_NoCoaxialFlow} (b). As indicated in \cref{fig:NeutralCurve_NoCoaxialFlow} (g-h), these two stationary modes $S_1$ and $S_2$ are localised within the recirculation bubble. For jet distances $L < 2$, the second mode that bifurcates is $F_1$ mode, depicted in  \cref{fig:NeutralCurve_NoCoaxialFlow} (i). This situation corresponds to a bifurcation scenario similar to other axisymmetric flow configurations, such as the flow past a sphere or a disk \cite{Auguste10,meliga2009global}. For larger distances between jets, the scenario changes. The second bifurcation from the axisymmetric steady-state is the $F_2$, displayed in \cref{fig:NeutralCurve_NoCoaxialFlow} (j). Other configurations where the primary or secondary instability involves modes with azimuthal component $m=2$ are swirling jets \cite{meliga2012weakly} and the wake flow past a rotating sphere \cite{sierra2022Rotating}. The unsteady modes $F_1$ and $F_2$ possess a much larger spatial support than $S_1$ and $S_2$. They are formed by an array of counter-rotating vortex spirals developing in the wake of the separating duct wall. For the mode $F_2$ the amplitude of these structures grows downstream of the nozzle, in the axial direction, with a maximum around $z\approx 10$, after which they slowly decay. The mode $F_1$ grows further downstream, with a maximum around $z \approx 50$. The spatial structure of these eigenmodes resembles the axisymmetric mode of Figure 9 in \cite{Cantonetal2017}. Thus, the steady modes and  unsteady modes differ in their spatial support, that is, even though both steady and unsteady modes are localised in space, the support of the steady ones is confined within the recirculation bubble. Instead, the unsteady modes are convected far downstream until they reach a maximum. This latter characteristic is classical of modes with a large transient growth, as it was noticed by \cite{Cantonetal2017}. On the other hand, the nature of the steady modes is similar to the symmetry-breaking instabilities behind the disk or the sphere. These modes are far less sensitive to transient growth and are observable with direct numerical simulations and experiments.

\subsection{Fixed distance between jets and variable velocity ratio \texorpdfstring{$\delta_u$}{du} \label{sec:resultsFD}}

\begin{figure}
    \centering
    \subfloat[]{\includegraphics[width=0.45\columnwidth]{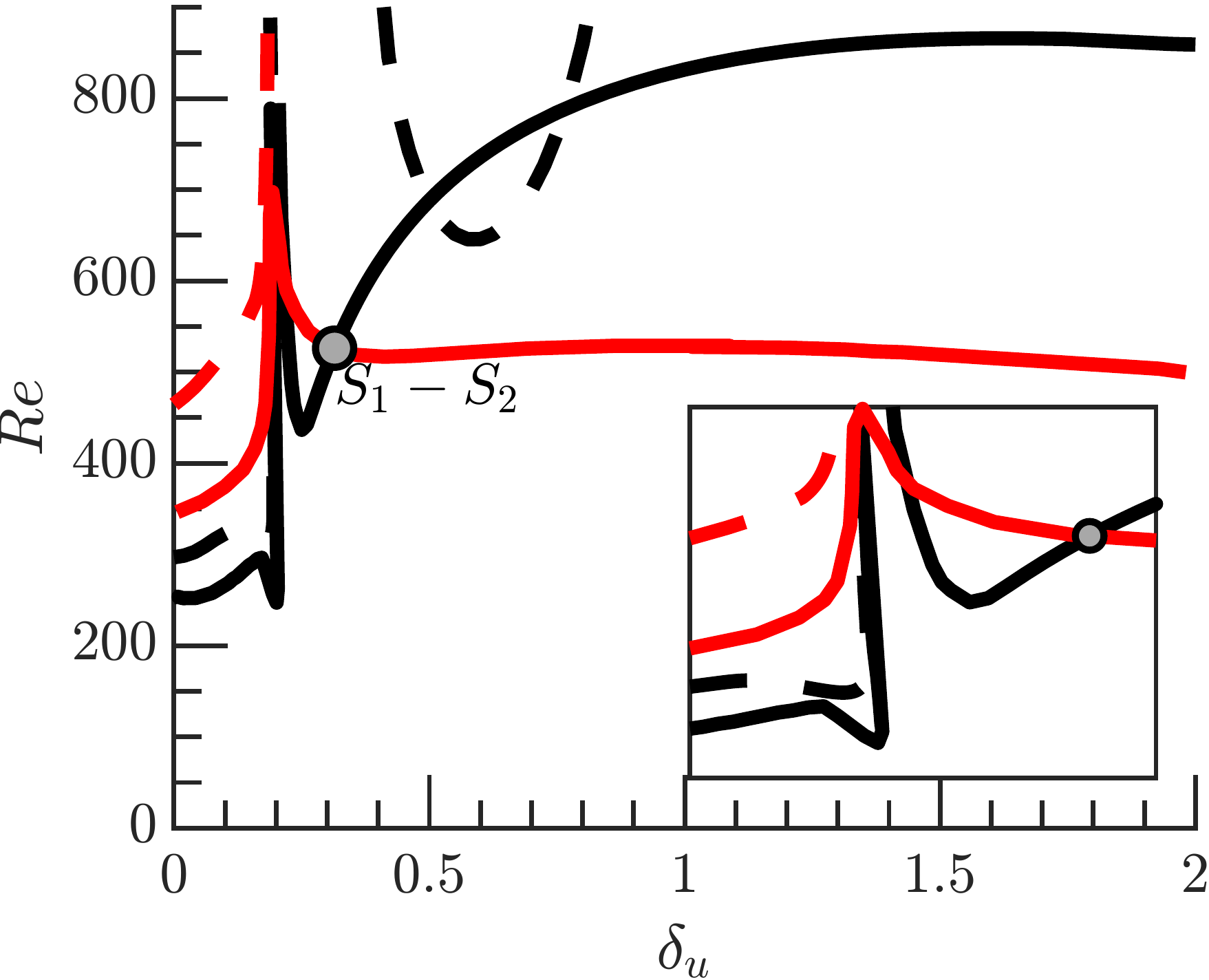}} 
    \subfloat[]{\includegraphics[width=0.45\columnwidth]{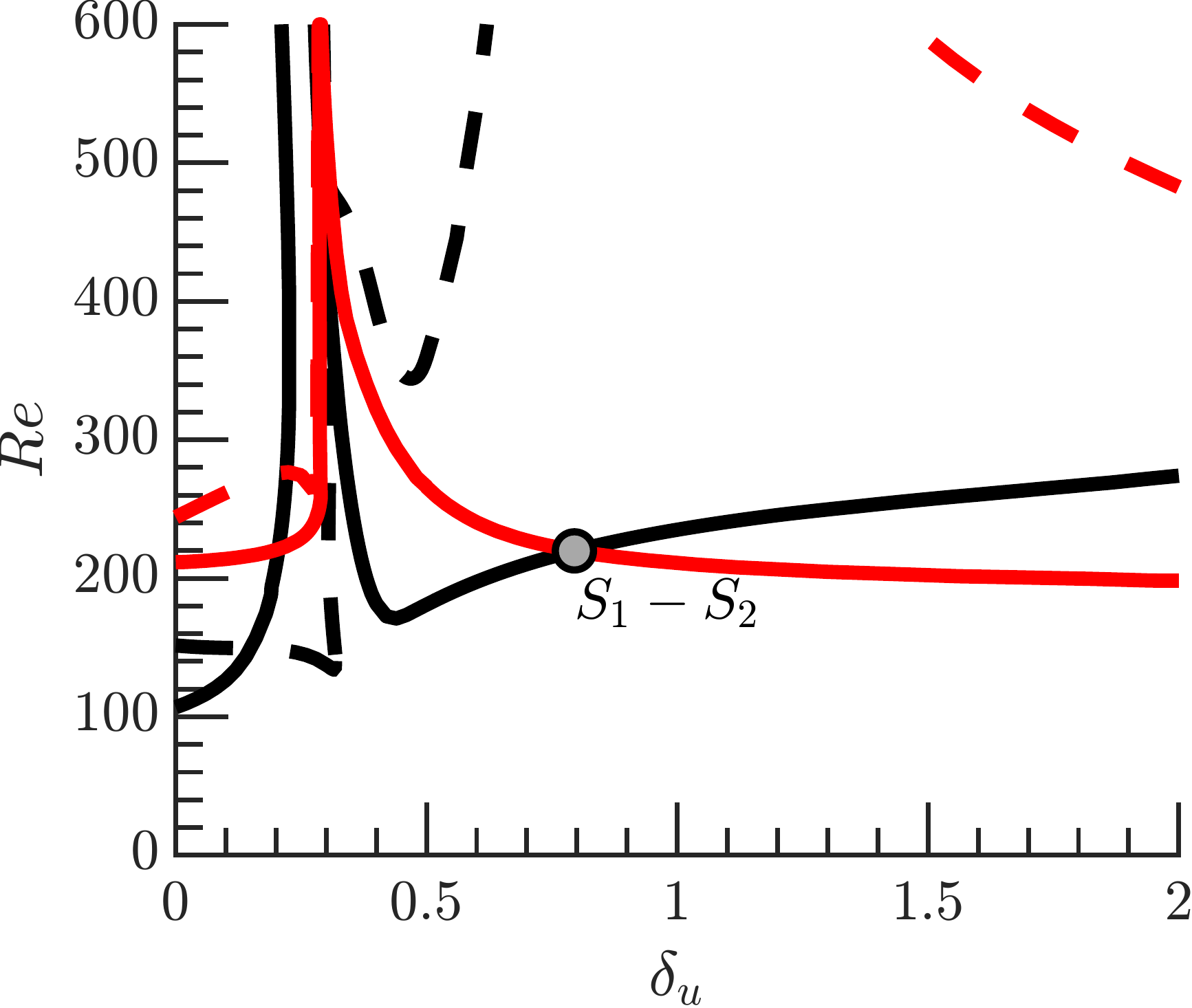}} \\
    \subfloat[]{\includegraphics[width=0.45\columnwidth]{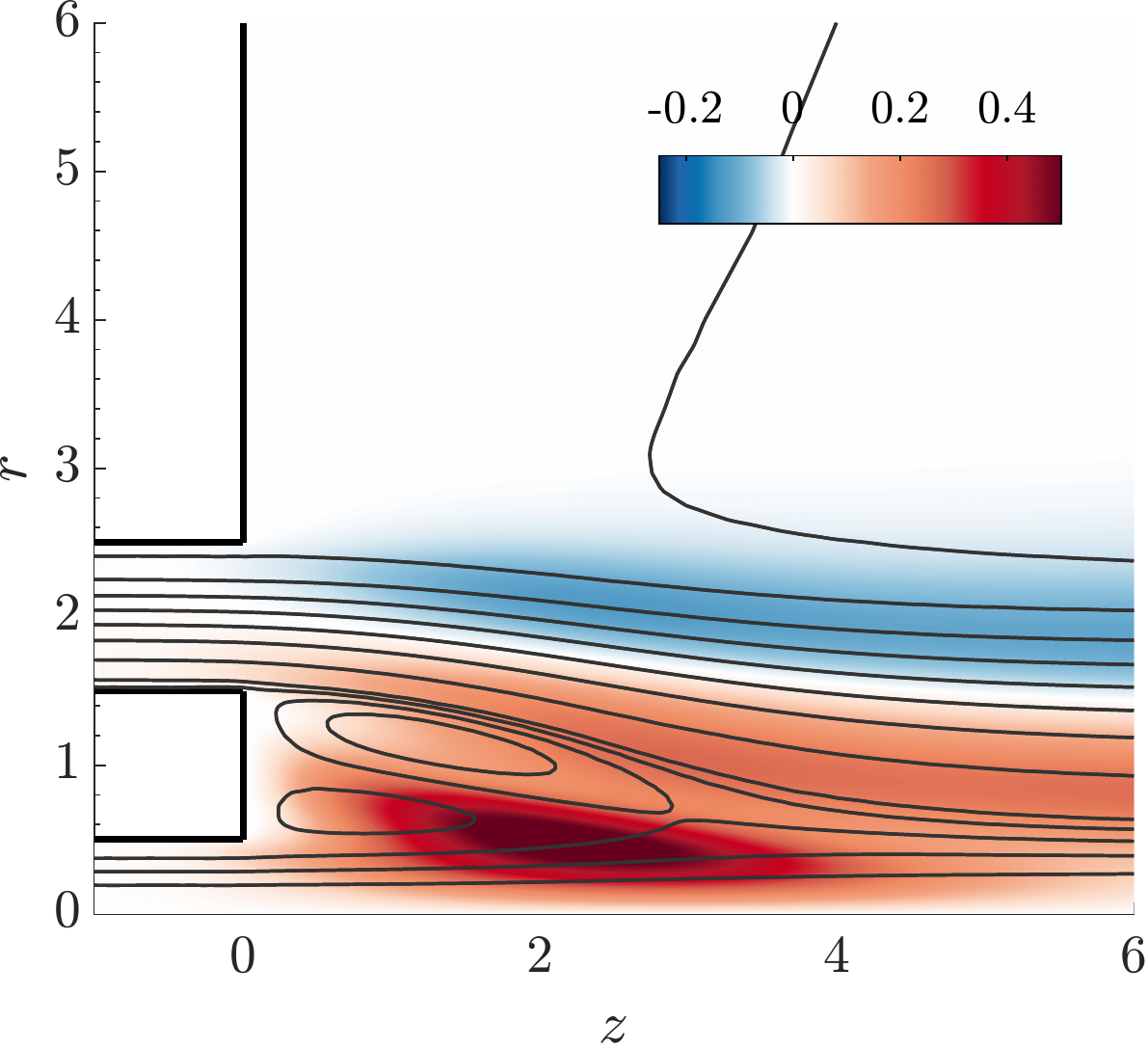}}     \subfloat[]{\includegraphics[width=0.45\columnwidth]{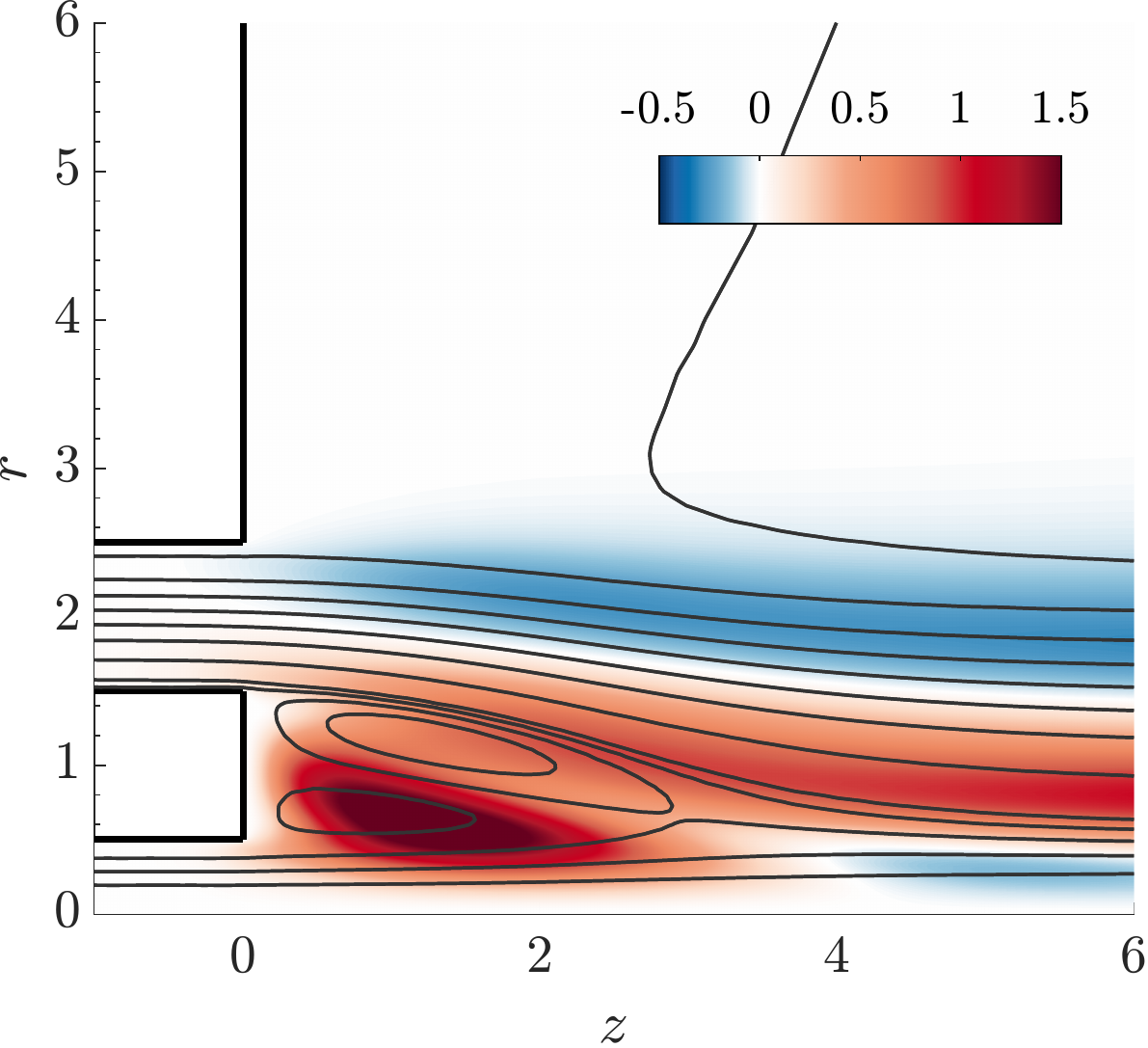}}
    \caption{ Linear stability boundaries for the concentric jets (a) $L = 0.5$ and (b) $L = 1$. Same legend as \cref{fig:NeutralCurve_NoCoaxialFlow}. Visualizations of real part of the streamwise axial velocity of the critical modes (c) $S_1$ and (d) $S_2$. }
    \label{fig:NeutralCurve_L0p5}
\end{figure}

In the following, we focus on the influence of the velocity ratio $\delta_u$ between jets for fixed jet distances $L$. \Cref{fig:NeutralCurve_L0p5} displays the neutral curve of stability for jet distances (a) $L=0.5$ and (b) $L=1$. One may observe that the primary bifurcation is not always associated to the mode $S_1$ as it is the case for $\delta_u=0$. For sufficiently large velocity ratios, the primary instability leads to a non-axisymmetric steady-state with a double helix. Another interesting feature, which could motivate a control strategy, is the occurrence of vertical asymptotes. This sudden change in the critical Reynolds number is due to the retraction and disappearance of the recirculating region. For $L=0.5$, this sudden change occurs for $\delta_u \approx 0.25$, and for higher values of $\delta_u$ the critical Reynolds number is around twice larger than the one of the annular jet ($\delta_u=0$). The case of jet distance $L=1$ was discussed in \cref{sec:SteadyState}. The sudden change in the stability of the branch $S_1$ occurs between $\delta_u \in [0.25,0.5]$. Within this narrow interval, the primary branch of instability is the $F_1$. At around $\delta_u=0.4$, the primary bifurcation is again the branch $F_1$, which becomes secondary at around $\delta_u \approx 0.8$ in favour of the branch $S_2$. In \cref{fig:NeutralCurve_L0p5}
we have highlighted the codimension two point interaction between the $S_1-S_2$ modes, whose modes are depicted in \cref{fig:NeutralCurve_L0p5} (c-d), which will be analysed in detail in \cref{sec:ModeInt}. Around this point, we can observe the largest stabilisation ratio between the annular jet ($\delta_u=0$) and a configuration of concentric jets ($\delta_u \neq 0$). 

\subsection{Fixed velocity ratio \texorpdfstring{$\delta_u$}{du} and variable distance between jets \label{sec:resultsFV}}

\begin{figure}
    \centering
    a)\hspace{4cm} b)\\
    \includegraphics[width=0.45\columnwidth]{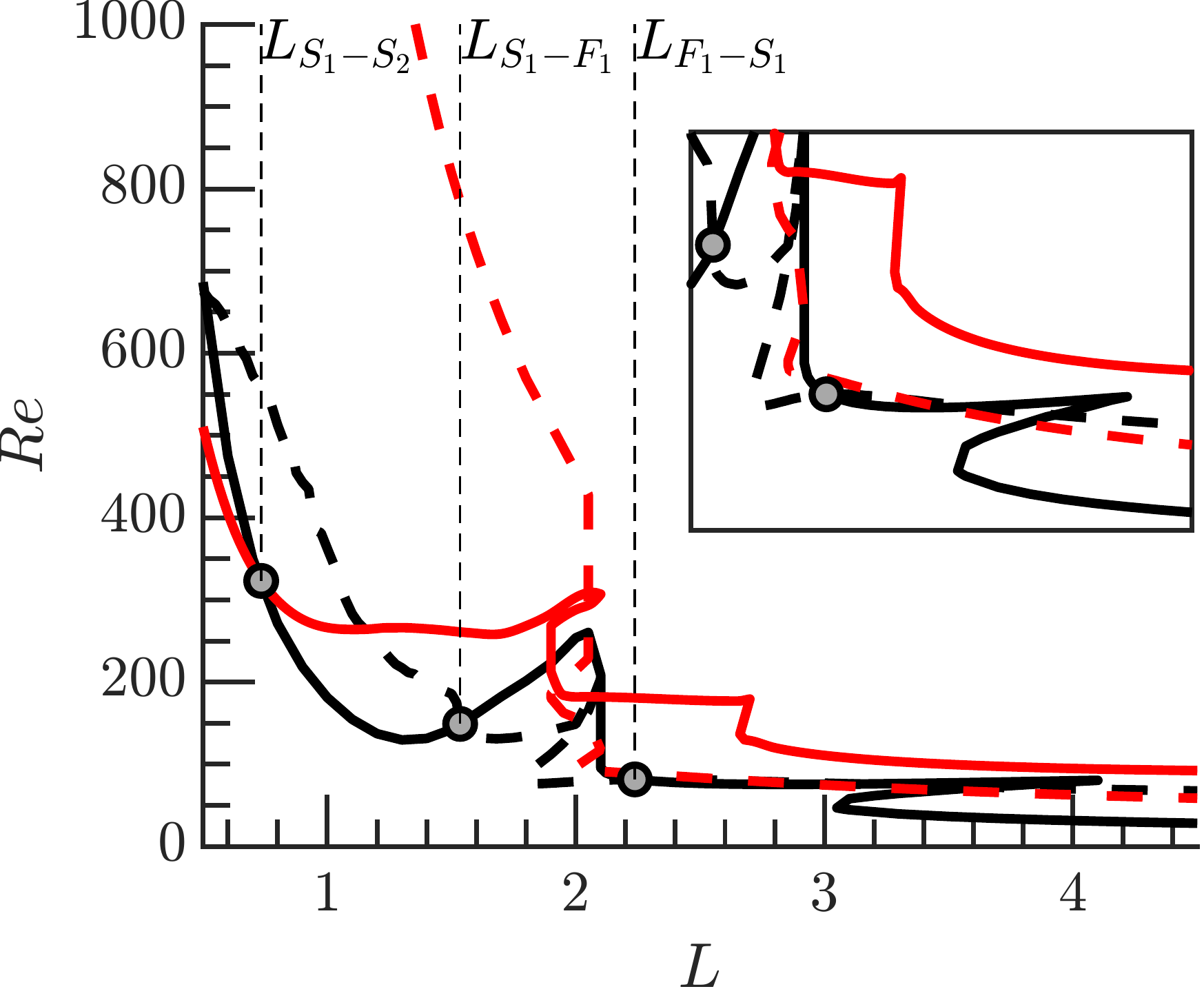}
    \includegraphics[width=0.45\columnwidth]{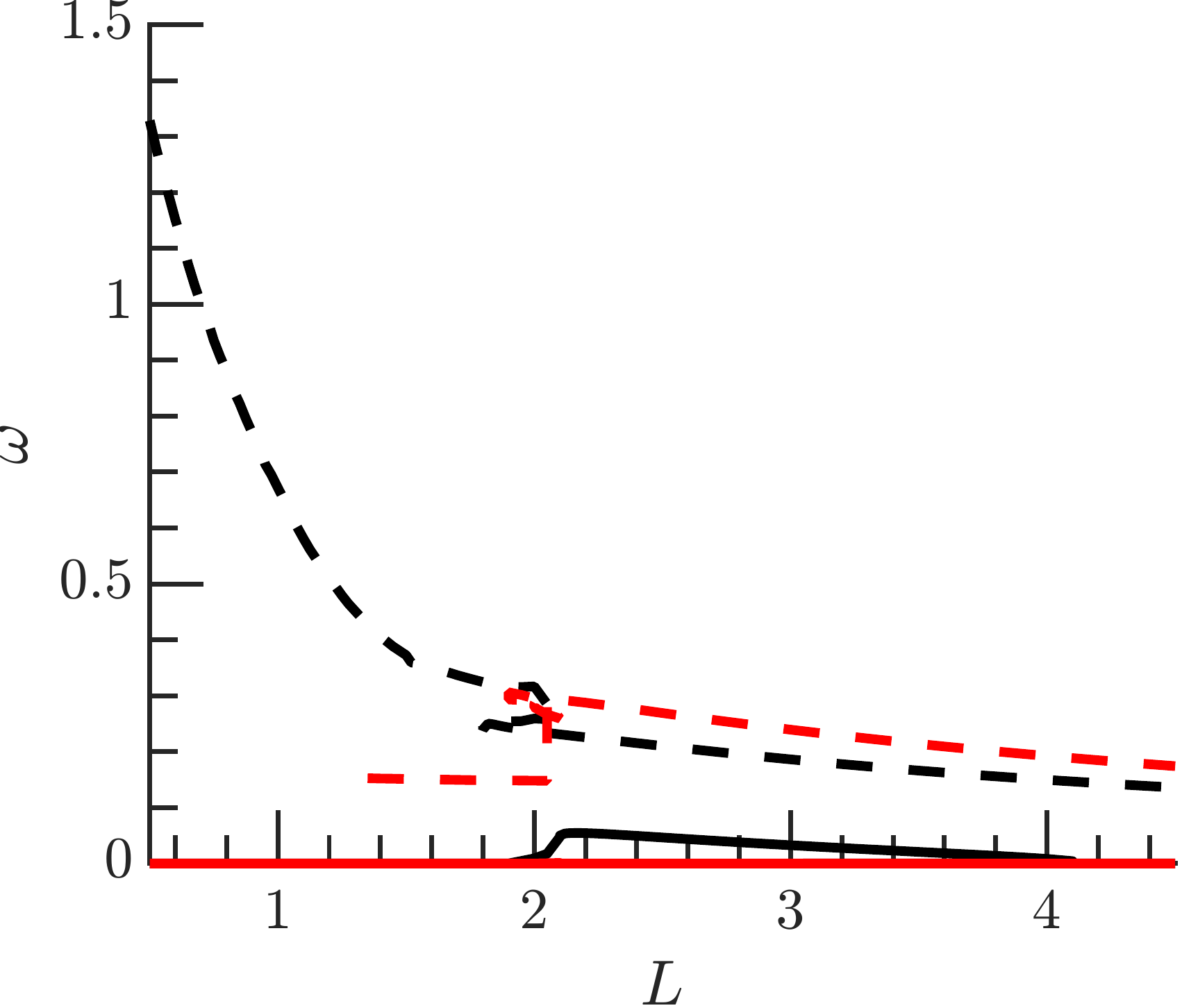}\\
    c)\hspace{4cm} d)\\
    \includegraphics[width=0.45\columnwidth]{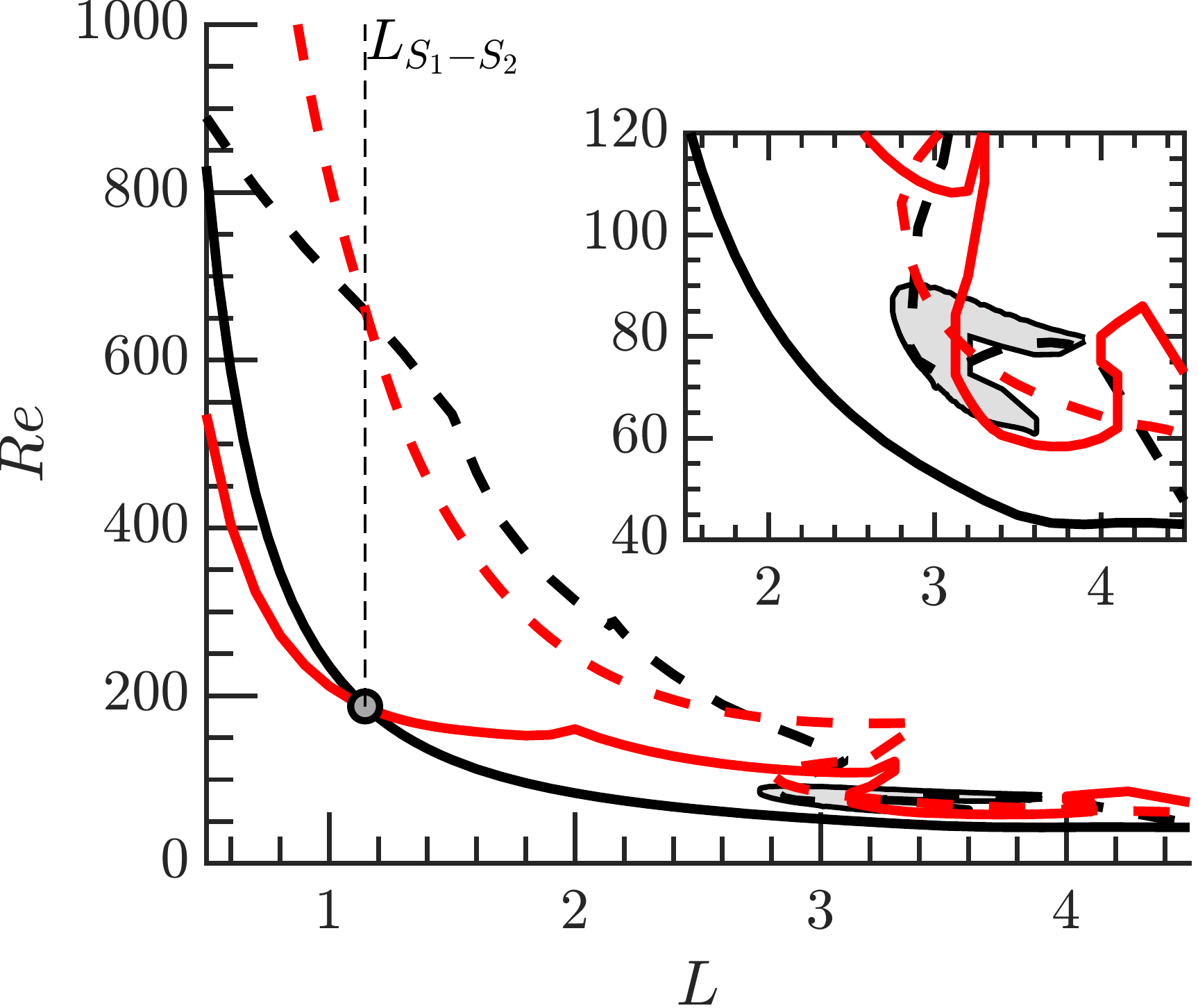}
    \includegraphics[width=0.45\columnwidth]{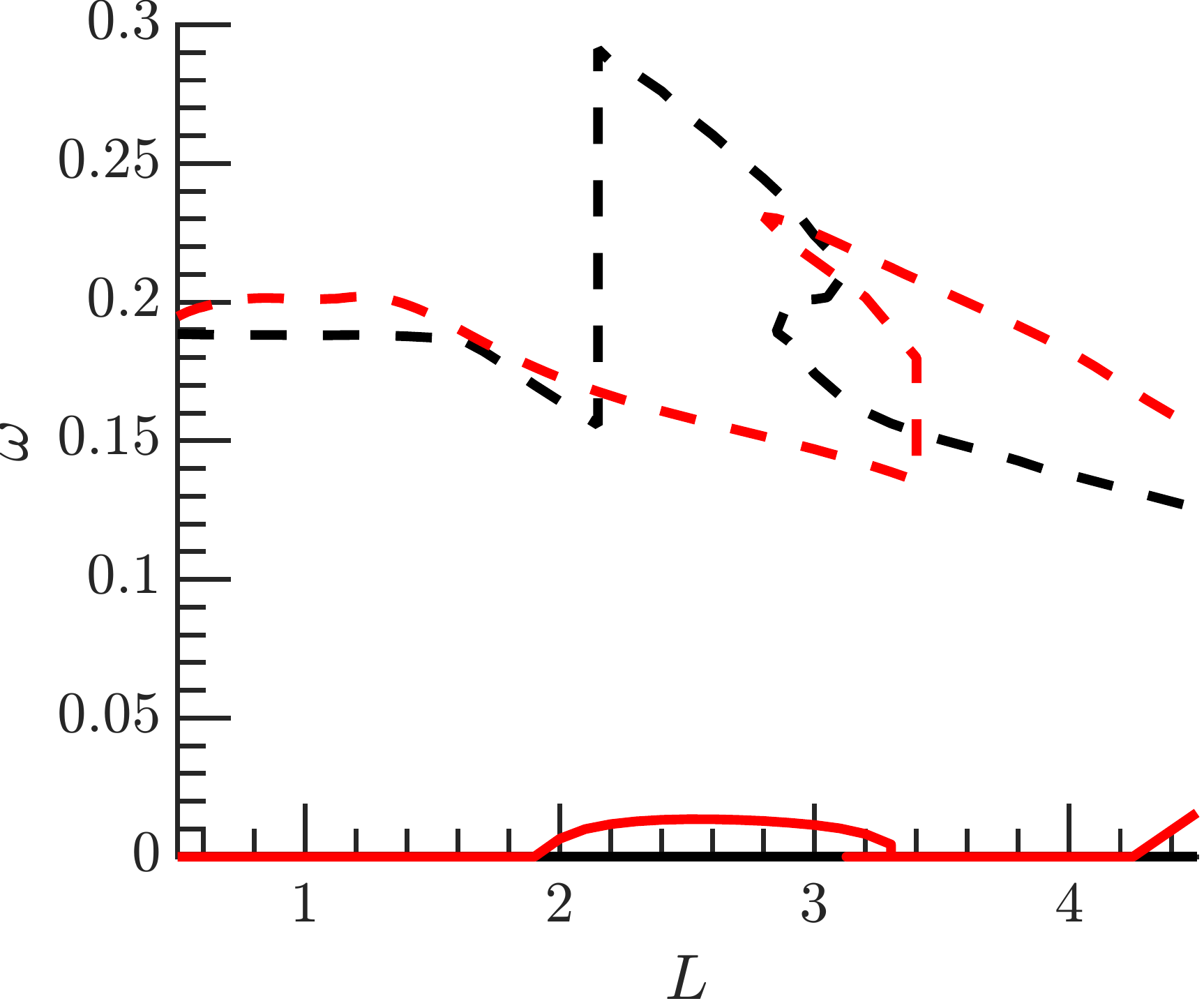}\\
    e)\hspace{4cm} f)\\
    \includegraphics[width=0.45\columnwidth]{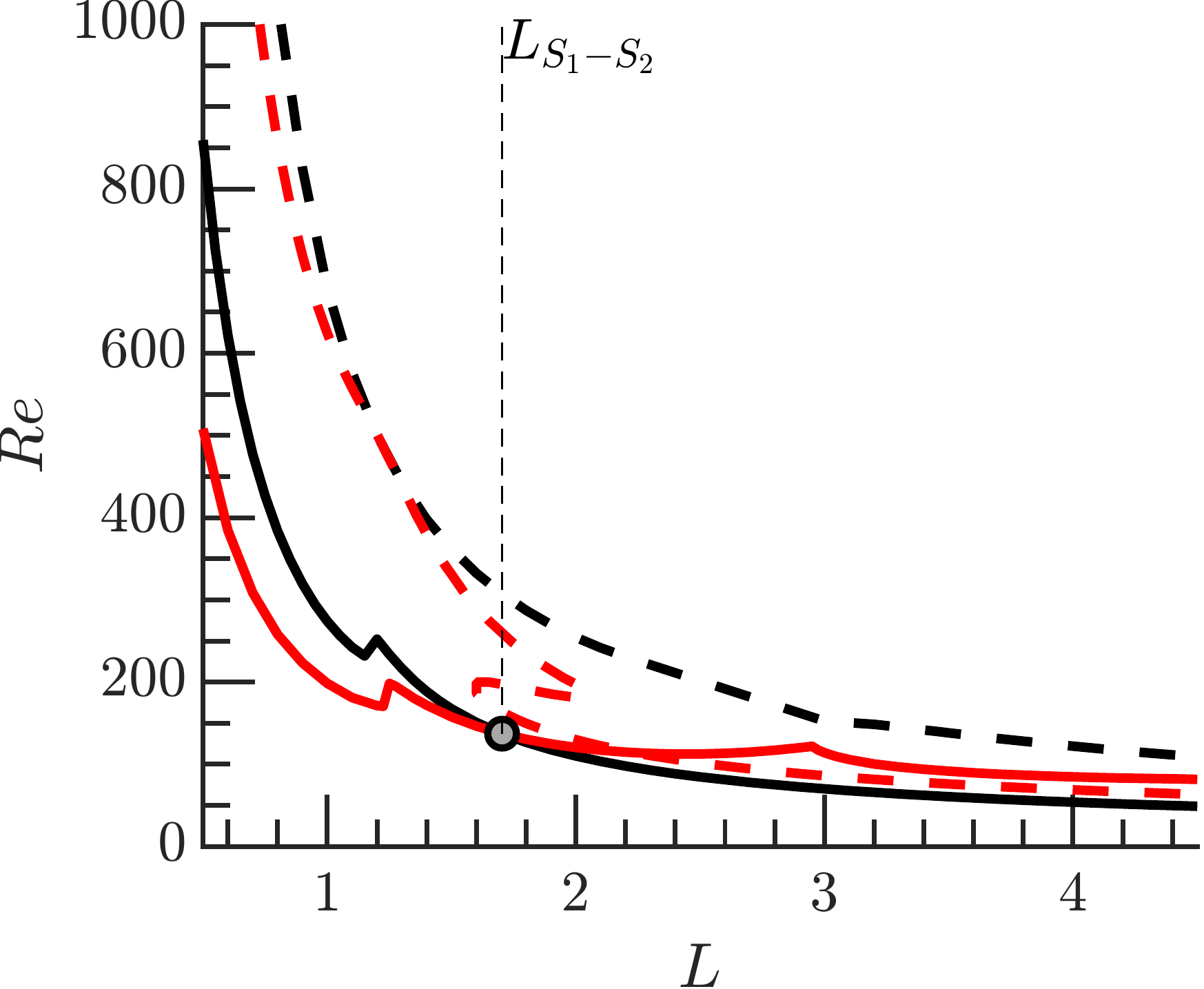}
    \includegraphics[width=0.45\columnwidth]{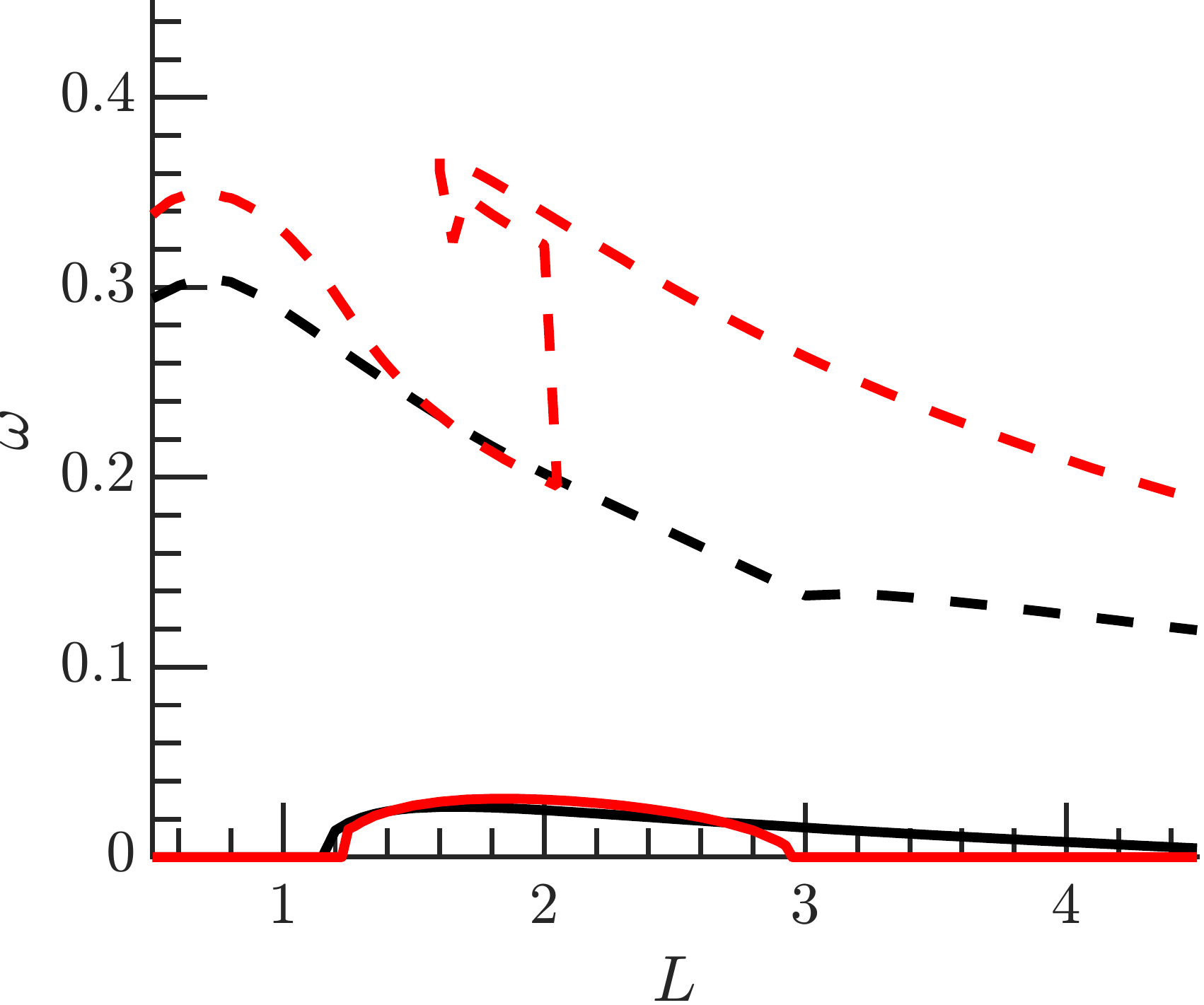}
    \caption{Neutral lines of the four modes found studying the configuration of two concentric jets fixing the velocity ratio. (a-b) $\delta_u =0.5$, (c-d) $\delta_u = 1$, (e-f) $\delta_u = 2$. Black lines: modes with $m=1$, red lines: modes with $m=2$. Straight lines: steady modes, dashed lines: unsteady modes.}
    \label{fig:NL_delta}
\end{figure}

\Cref{fig:NL_delta} compares the results obtained for a constant velocity ratio when varying the distance between jets. As observed before, the solution becomes more unstable by increasing the distance between jets. The largest critical Reynolds number is found at $\delta_u=0$. The critical Reynolds decreases with the distance between jets $L$. The points of codimension two, i.e., the points where mode switching occurs, are highlighted in \cref{fig:NL_delta}. We can appreciate that the interaction between the branch $S_1$ and $S_2$ happens for every velocity ratio $\delta_u$ explored, and it is the mode interaction associated to the smallest distance between jets. Additionally, for a velocity ratio $\delta_u=0.5$ there exists two points where the branches of the linear modes $S_1$ and $F_1$ intersect. 
Another feature of the neutral curves is the existence of turning points, which are associated to the existence of saddle node bifurcations of the axisymmetric steady-state. The saddle node bifurcations of the steady-state induces the existence of regions in the neutral curves with a \textit{tongue} shape. These saddle node bifurcations are also responsible for the formation of the vertical asymptotes observed in \cref{fig:NeutralCurve_L0p5}. 
Finally, it is of interest the transition of the modes $S_1$ and $S_2$, which induce the symmetry breaking of the axisymmetric steady state to slow low frequency spiralling structures. These modes have been identified for $\delta_u=0.5$ for $m=1$, $\delta_u=1$ for $m=2$, and $\delta_u=2$ for both $m=1$ and $m=2$. As it will be clarified in \cref{sec:ModeInt}, these oscillations are issued from the non-linear interaction of modes, emerging simultaneously for a specific Reynolds number, and changing their position as the most unstable global mode of the flow. 

\section{Mode interaction between two steady states. Resonance \texorpdfstring{$1:2$}{1:2} \label{sec:ModeInt}}
\subsection{Normal form, basic solutions and their properties}

The linear diagrams of \cref{sec:LST} have shown the existence of the mode interaction between the modes $S_1$ and $S_2$. It corresponds roughly to the mode interaction that occurs at the largest critical Reynolds number for any value of $L$ herein explored. 
In this section, we analyse the dynamics near the $S_1:S_2$ organizing centre. We perform a normal form reduction, which allows us to predict non-axisymmetric steady, periodic, quasiperiodic and heteroclinic cycles between non-axisymmetric states.  \\
The mode interaction that is herein analysed corresponds to a steady-steady bifurcation with $O(2)$ symmetry and with strong resonance $1:2$. Such a bifurcation scenario has been extensively studied in the past by \cite{dangelmayr1986steady,jones1987strong,porter2001new,armbruster1988heteroclinic} and the reflection symmetry breaking case (SO(2)) by \cite{porter2005dynamics}. 
In order to unravel the existence and the stability of the nonlinear states near  the codimension two point, let write the flow field as
\bes{Normal_Form_AnsatzSS}
\displaystyle \vec{q} &= \QVBF + \text{Re} \big[ r_1(\tau) e^{i \phi_1(\tau)} e^{-i\theta} \hat{\vec{q}}_{s,1}  \big] + \text{Re} \big[ r_2(\tau) e^{i \phi_2(\tau)} e^{-2i \theta} \hat{\vec{q}}_{s,2}  \big] \\
\ees
in polar coordinates for the complex amplitudes $z_1 = r_1 \text{e}^{\im \phi_1}$ and $z_2 = r_2 \text{e}^{\im \phi_2}$ where $r_j$ and $\phi_j$ for $j=1,2$ are the amplitude and phase of the symmetry-breaking modes $m=1$ and $m=2$, respectively. The complex-amplitude normal form \cref{eq:dynsyst} is expressed in this reduced polar notation as follows,
\begin{subequations}
\begin{align}
\besp
\dot{r}_{1} = e_3 r_{1} r_{2} \cos(\chi) + r_{1} \Big( \lambda_{(s,1)} + c_{(1,1)} r_{1}^2 + c_{(1,2)} r_{2}^2 \Big),
\eesp \\
\besp
\dot{r}_{2} = e_4 r_{1}^2 \cos(\chi) + r_{2} \Big( \lambda_{(s,2)} + c_{(2,1)} r_{1}^2 + c_{(2,2)} r_{2}^2 \Big),
\eesp \\
\besp
\dot{\chi} = - \Big( 2 e_3 r_{2} + e_4 \frac{r_{1}^2}{r_{2}}  \Big) \sin(\chi),
\eesp
\end{align}
\label{eq:NormalFormThirdOrderSteadySteady}
\end{subequations}
where the phase $\chi = \phi_2 - 2\phi_1$ is coupled with the amplitudes $r_1$ and $r_2$ because of the existence of the $1:2$ resonance. The individual phases evolve as
\bes{PhasesNormalForm}
    \dot{\phi}_1 &= e_3 r_2 \sin(\chi), \\
    \dot{\phi}_2 &= - e_4 \frac{r_1^2}{r_2} \sin(\chi).
\ees
Before proceeding to the analysis of the basic solutions of \cref{eq:NormalFormThirdOrderSteadySteady}, we can simplify these equations by the rescaling 
$$ \big( \frac{r_1}{|e_3 e_4|^{1/2}}, \frac{r_2}{e_3} \big) \to (r_1,r_2),  $$
which yields the following equivalent system 
\begin{subequations}
\begin{align}
\besp
\dot{r}_{1} = r_{1} r_{2} \cos(\chi) + r_{1} \Big( \lambda_{(s,1)} + c_{11} r_{1}^2 + c_{12} r_{2}^2 \Big),
\eesp \\
\besp
\dot{r}_{2} = s r_{1}^2 \cos(\chi) + r_{2} \Big( \lambda_{(s,2)} + c_{21} r_{1}^2 + c_{22} r_{2}^2 \Big),
\eesp \\
\besp
\dot{\chi} = - \frac{1}{r_2} \Big( 2  r_{2}^2 + s r_1^2  \Big) \sin(\chi), 
\eesp
\end{align}
\label{eq:NormalFormThirdOrderSteadySteadyRescaling}
\end{subequations}
where the coefficients 
$$s = \text{sign}(e_3 e_4), \quad c_{11} = \frac{c_{(1,1)}}{|e_3 e_4|}, \quad c_{12} = \frac{c_{(1,2)}}{e_3^2}, \quad c_{21} = \frac{c_{(2,1)}}{|e_3 e_4|}, \quad c_{22} = \frac{c_{(2,2)}}{e_3^2}.$$ Finally, we consider a third normal form equivalent to the previous ones but which removes the singularity of \cref{eq:NormalFormThirdOrderSteadySteady,eq:NormalFormThirdOrderSteadySteadyRescaling} when $r_2 = 0$. Standing waves ($\sin{\chi} = 0$) naturally encounter this type of artificial singularity, which manifests as in \cref{eq:NormalFormThirdOrderSteadySteadyRescaling} as an instantaneous jump from one standing subspace to the other by a $\pi$-translation. This is the case of the heteroclinic cycles, previously studied by \cite{armbruster1988heteroclinic,porter2001new}. The third normal form, which we shall refer to as reduced Cartesian normal form, takes advantage of the simple transformation $x=r_2 \cos(\chi)$, $y=r_2 \sin(\chi)$ \cite{porter2005dynamics}:
\begin{subequations}
\begin{align}
\besp
\dot{r}_{1} = r_{1} \Big( \lambda_{(s,1)} + c_{11} r_{1}^2 + c_{12} (x^2+y^2) + x \Big),
\eesp \\
\besp
\dot{x} = s r_{1}^2 + 2 y^2 +  x\Big( \lambda_{(s,2)} + c_{21} r_{1}^2 + c_{22} (x^2+y^2) \Big),
\eesp \\
\besp
\dot{y} = - 2 xy +  y \Big( \lambda_{(s,2)} + c_{21} r_{1}^2 + c_{22} (x^2+y^2) \Big),
\eesp
\end{align}
\label{eq:NormalFormThirdOrderSteadySteadyCartesian}
\end{subequations}
In this final representation standing wave solutions are contained within the invariant plane $y=0$, and due to the invariance of \cref{eq:NormalFormThirdOrderSteadySteadyCartesian} under the reflection $y \mapsto -y$, one can restrict attention, without loss of generality, to solutions with $y \geq 0$, cf~\cite{porter2001new}. 

The system \cref{eq:NormalFormThirdOrderSteadySteadyRescaling} possess four types of fixed points, which are listed in \cref{tab:DefinitionModes}. 

\begin{table*}
\begin{tabular}{llll}
\hline
\textrm{Name}&
\textrm{Definition}&
\textrm{Bifurcations}&
\textrm{Comments} \\
\hline
 O & $r_{1,O}=r_{2,O}=0$ & $-$ & Steady axisymmetric state \\
\hline
 P & $r_{2,P}^2 = \frac{-\lambda_{(s,2)}}{c_{22}}$, $r_{1,P}=0$ & $\lambda_{(s,2)} = 0$ & Bifurcation from O \\
\hline
 & $r_{1,MM} = - \frac{ \lambda_{(s,1)} \pm  r_{2,MM} + c_{12} r_{2,MM}^2 }{c_{11}}$ & $\lambda_{(s,1)} = 0$ & Bifurcation from O \\
MM & $ \text{P}_{\text{MM}}(r_{2,MM} \cos(\chi_{MM})) = 0 $ & $\sigma_{\pm} = 0$ & Bifurcation from P \\
& $\cos(\chi_{MM}) = \pm 1$ &  & \\
\hline
& $\cos(\chi_{TW}) = \frac{(2 c_{11} + c_{12}) \lambda_{(s,2)} - (2 c_{21} + c_{22}) \lambda_{(s,1)} }{ \Sigma_{TW} ( 2 \lambda_{(s,1)} + \lambda_{(s,2)})  }$  & & \\
TW & $r_{2,TW}^2 = \frac{ -( 2 \lambda_{(s,1)} + \lambda_{(s,2)})  }{\Sigma_{TW}} $  & $\cos(\chi_{TW}) = \pm 1 $ & Bifurcation from MM \\
 & $r_{1,TW}^2 = 2 r^2_{2,TW}$ &  & \\
\hline
\end{tabular}
\caption{\label{tab:DefinitionModes}%
Definition of the fixed points of the reduced polar normal form \cref{eq:NormalFormThirdOrderSteadySteadyRescaling}. $\sigma_{\pm}$ is defined in \cref{eq:SigmaPP}, the polynomial $P_{MM}$ is defined in \cref{eq:PMM} and $\Sigma_{TW} \equiv 4 c_{11} + 2 (c_{12} + c_{21}) + c_{22}$. }
\end{table*}

First, the axisymmetric steady state (O) is represented by $(r_1,r_2) = (0,0)$, so it is the trivial steady-state of the normal form. The second steady-state is what it is denoted as pure mode (P). In the original coordinates, it corresponds to the symmetry breaking structure associated to the mode $S_2$. This state bifurcates from the axisymmetric steady state (O) when $\lambda_{(s,2)} = 0$. The third fixed point is the mixed mode state (MM), which is listed in \cref{tab:DefinitionModes}. It corresponds to the reflection symmetry preserving state associated to the mode $S_1$. It may bifurcate directly from the trivial steady state O, when $\lambda_{(s,1)} = 0$ or from P whenever $\sigma_{+} = 0$ or $\sigma_{-} = 0$, where $\sigma_{\pm}$ is defined as 
\be{SigmaPP}
\sigma_{\pm} \equiv \lambda_{(s,1)} -  \frac{-\lambda_{(s,2)} c_{12}}{c_{22}} \pm \sqrt{\frac{-\lambda_{(s,2)}}{c_{22}}} . 
\ee
 The representation in the reduced polar form is 
$$ r_{1,MM} = - \frac{ \lambda_{(s,1)} \pm r_{2,MM} + c_{12} r_{2,MM}^2 }{c_{11}}, \qquad \cos(\chi_{MM}) = \pm 1, $$
and the condition $\text{P}_{\text{MM}}(r_{2,MM} \cos(\chi_{MM})) = 0$, where  $\text{P}_{\text{MM}}$ is defined as 
\be{PMM}
 \text{P}_{\text{MM}}(x) \equiv s \mu_1 + (s + c_{21} \lambda_{(s,1)} - c_{(1,1)} \lambda_{(s,2)}) x + (c_{21} + s c_{12}) x^2 + (c_{12}c_{21} - c_{11}c_{22})x^3.  
\ee
Finally, the fourth fixed point of the system are travelling waves (TW). It is surprising that the interaction between two steady-states causes a time-periodic solution. The travelling wave emerges from MM in parity-breaking pitchfork bifurcation that breaks the reflection symmetry when $\cos(\chi_{TW}) = \pm 1 $. The TW drifts at a steady rotation rate  $\omega_{TW}$ along the group orbit, i.e., the phases $\dot{\phi}_1 = r_{2,TW} \sin(\chi_{TW})$ and $\dot{\phi}_2 = -s \frac{r_{1,TW}^2}{r_{2,TW}} \sin(\chi_{TW})$ are non-null. 

\begin{table*}
\begin{tabular}{lll}
\hline
\textrm{Name}&
\textrm{Bifurcation condition}&
\textrm{Comments} \\
\hline
 SW & $ s r_1^2 - 2c_{11} r_1^2 r_{2,MM} \cos(\chi_{MM}) - 2 c_{22} r_{2,MM}^3 \cos(\chi_{MM})^3 = 0 $ & Bif. from MM \\
 \hline
 MTW & $D_{TW} - T_{TW} I_{TW} = 0$, $I_{TW} > 0$ & Bif. from TW \\
\hline
\end{tabular}
\caption{\label{tab:DefinitionModesLC}%
Definition of the limit cycles of the reduced polar normal form \cref{eq:NormalFormThirdOrderSteadySteadyRescaling}. }
\end{table*}

Mixed modes and travelling waves may further bifurcate into standing waves (SW) and modulated travelling waves (MTW), respectively. These are generic features of the $1:2$ resonance for small values of $\lambda_{(s,1)}$ and $\lambda_{(s,2)}$, when $s = -1$. In the original coordinates, SW are periodic solutions, whereas MTW are quasiperiodic.
Standing waves emerge via a Hopf bifurcation from MM when the conditions $\text{P}_{\text{SW}} \Big(r_{2,MM} \cos(\chi_{MM}) \Big) > 0$ for $$ \text{P}_{\text{SW}} (x) \equiv (2 c_{22} x^3 - s r_1^2) c_{11} - (2 c_{12} x + 1)(c_{21}x + s)x, $$
and the one listed in \cref{tab:DefinitionModesLC} are satisfied. MTW are created when a torus bifurcation happens on the travelling wave branch when the conditions listed in \cref{tab:DefinitionModesLC} are satisfied. 

\begin{table*}
\begin{tabular}{lll}
\hline
\textrm{Name}&
\textrm{Condition}&
\textrm{Comments}\\
\hline
 Ht AGH & $\lambda_{(s,1)} > 0$, $\lambda_{(s,2)} > 0$, $c_{22} < 0$  & Existence\\
 & $ \sigma_{+} > 0, \sigma_{-} < 0$ & Asymptotic stability \\
\hline
\end{tabular}
\caption{\label{tab:DefinitionModesHt}%
Definition of the conditions for the existence of the Ht AGH (robust heteroclinic cycles connecting pure modes) of the reduced polar normal form \cref{eq:NormalFormThirdOrderSteadySteadyRescaling}. }
\end{table*}

\begin{figure}
    \centering
    {\includegraphics[width=0.45\columnwidth]{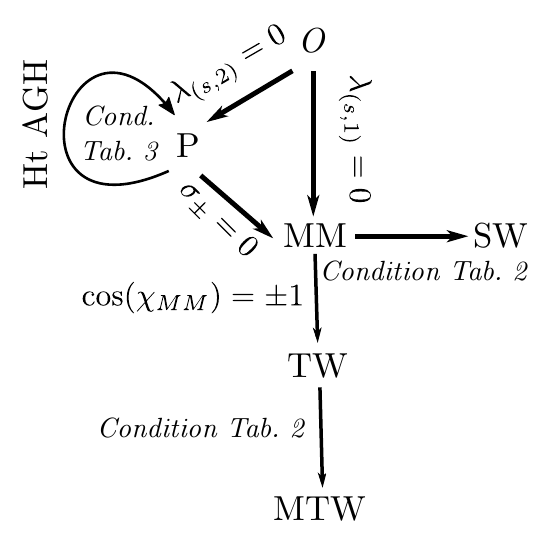}} 
    \caption{ Schematic representation of the basic solutions of \cref{eq:NormalFormThirdOrderSteadySteady} and their bifurcation path.}
    \label{fig:BifurcationDiagramSS}
\end{figure}

Another remarkable feature of \cref{eq:NormalFormThirdOrderSteadySteady} is the existence of robust heteroclinic cycles that are asymptotically stable. When $s = -1$, there are open sets of parameters (see \cref{tab:DefinitionModesHt})  where the reduced polar normal form exhibits structurally stable connections between $\pi-$translations on the circle of pure modes, cf~\cite{armbruster1988heteroclinic}. These structures are robust and have been observed in a large variety of systems, \cite{nore20031,nore2005experimental,mercader2002robust,palacios1997cellular,mariano2005computational}. In addition to these robust heteroclinic cycles connecting pure modes, there exists more complex limit cycles connecting O, P, MM and SW, cf~\cite{porter2001new}. These cycles are located for larger values of $\lambda_{(s,1)}$ and $\lambda_{(s,2)}$, with possibly chaotic dynamics (Shilnikov type). In this study, we have not identified any of these. Finally, a summary of the basic solutions and the bifurcation path is sketched in \cref{fig:BifurcationDiagramSS}.

\subsection{Results of the steady-steady $1:2$ mode interaction}
 \Cref{sec:resultsFV} reported the location of mode interaction points for discrete values of the velocity ratio $\delta_u$. The location of the mode interaction between $S_1$ and $S_2$ is depicted in \cref{fig:CodimensionTwoEvolution}. It shows that the mode switching between the modes $S_1$ and $S_2$ is indeed stationary only for $\delta_u < 1.5$ and $L < 1.3$. For larger values of the velocity ratio and the jet distance, the interaction is not purely stationary; at least one of the linear modes oscillates with a slow frequency. It implies that the mode selection for large velocity ratios near the codimension two points is similar to the one reported by \cite{meliga2012weakly} for swirling jets. However, even when the two primary bifurcations are non-oscillating ($S_1$ and $S_2$), the $1:2$ resonance of the azimuthal wavenumbers induces a slow frequency, what we denote as travelling wave solutions (TW). 

\begin{figure}
    \centering
    a)\hspace{4cm} b)\\
    \includegraphics[width=0.42\columnwidth]{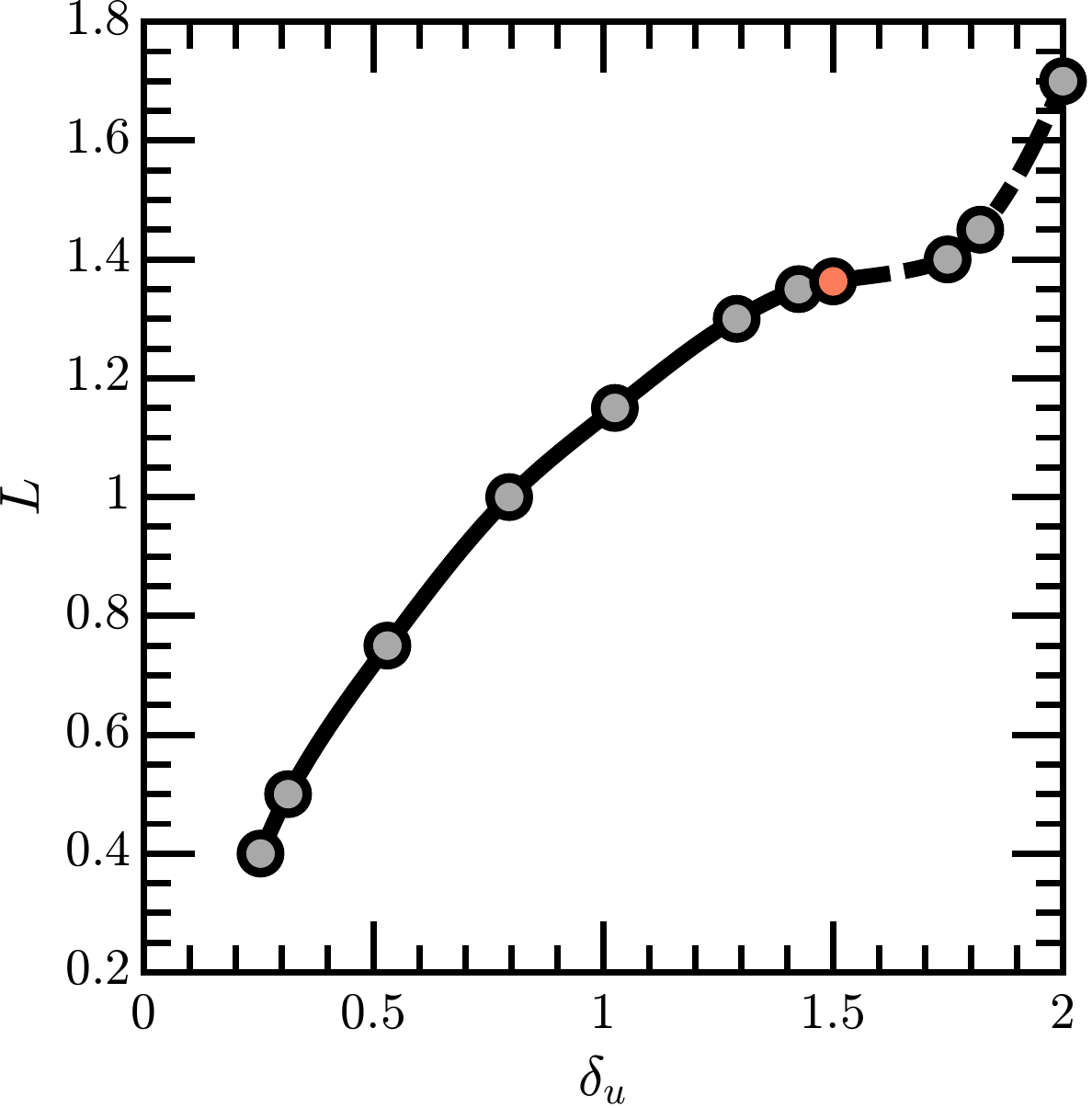} \quad
    \includegraphics[width=0.45\columnwidth]{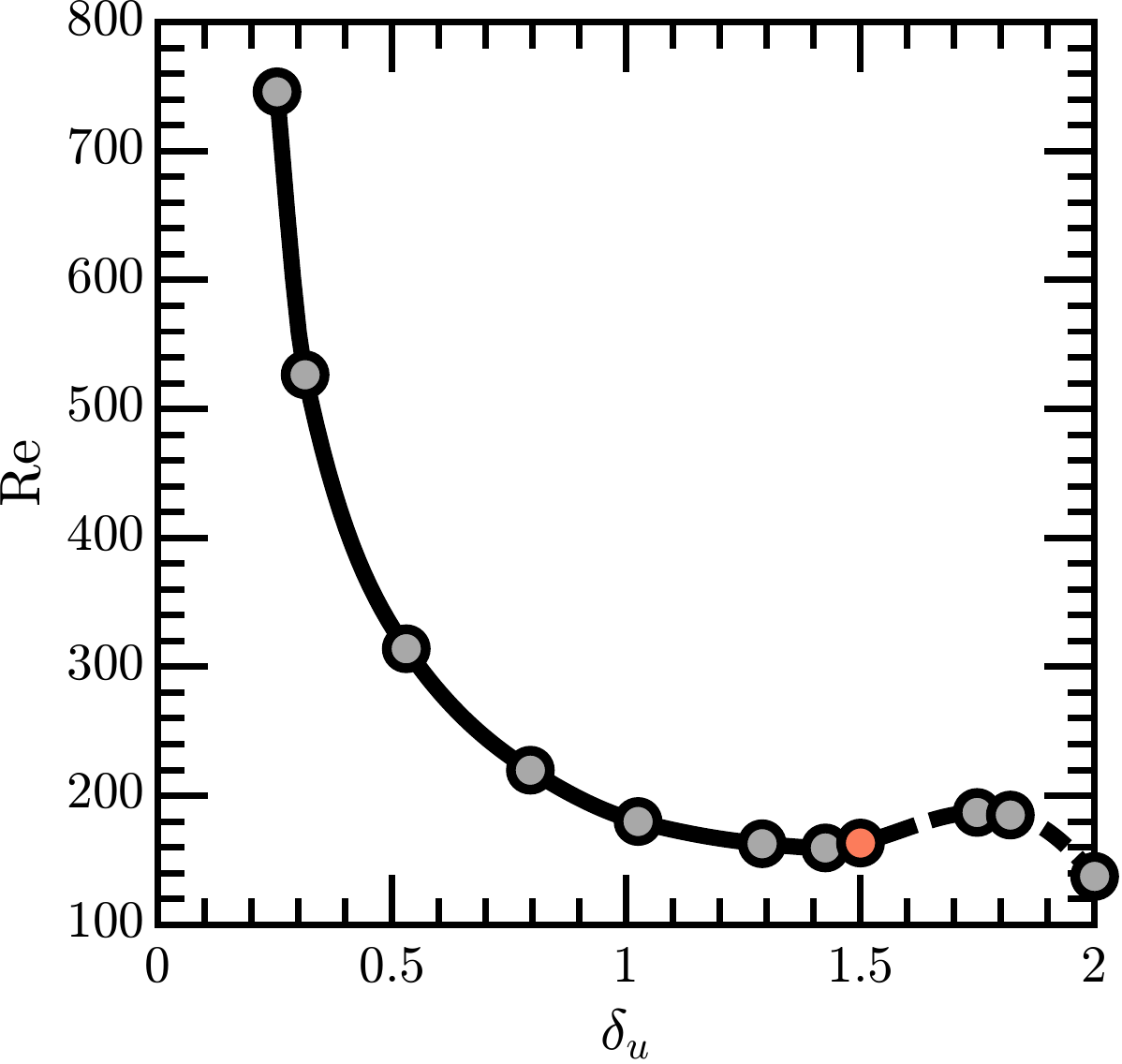}
    \caption{ Evolution of the codimension two interaction $S_1-S_2$ in the space of parameters $(\text{Re},L,\delta_u)$. Grey points denote the points that were computed and the red point denotes the transition from steady to unsteady with low frequency as reported in \cref{sec:resultsFV}. }
    \label{fig:CodimensionTwoEvolution}
\end{figure}

\begin{figure}
    \centering
    \includegraphics[width=1.0\columnwidth]{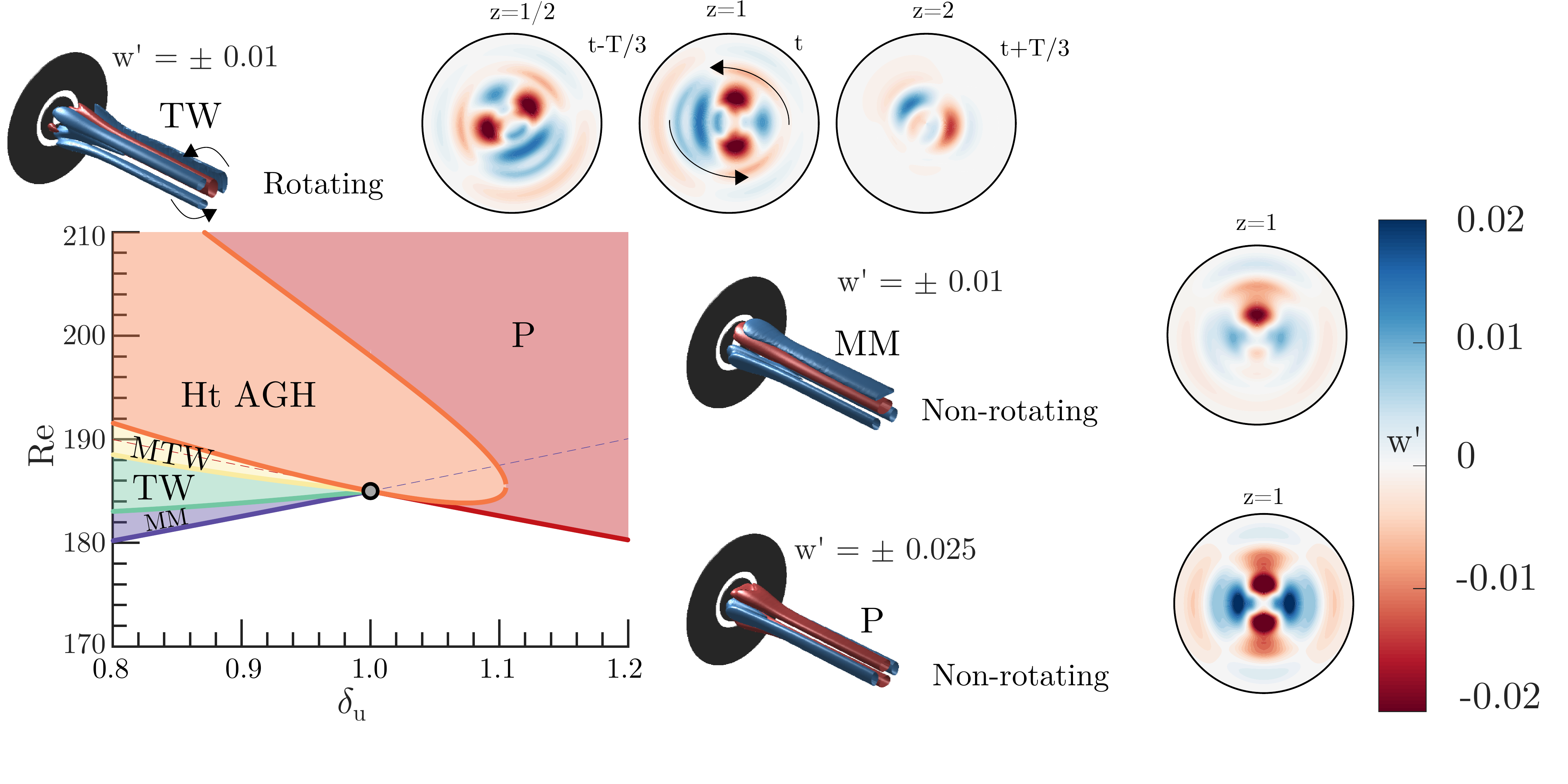} \quad
    \caption{ Phase portrait at the codimension two point $S_1:S_2$ for parameter values $(L,\delta_u) = (1.15,1.0)$. Visualisations of blue and red surfaces in the isometric views represent the respective positive and negative
isocontour values of the perturbative axial velocity indicated in the figure.  }
    \label{fig:PatternFormationSteadySteady}
\end{figure}

\begin{figure}
    \centering
    \includegraphics[width=0.4\columnwidth]{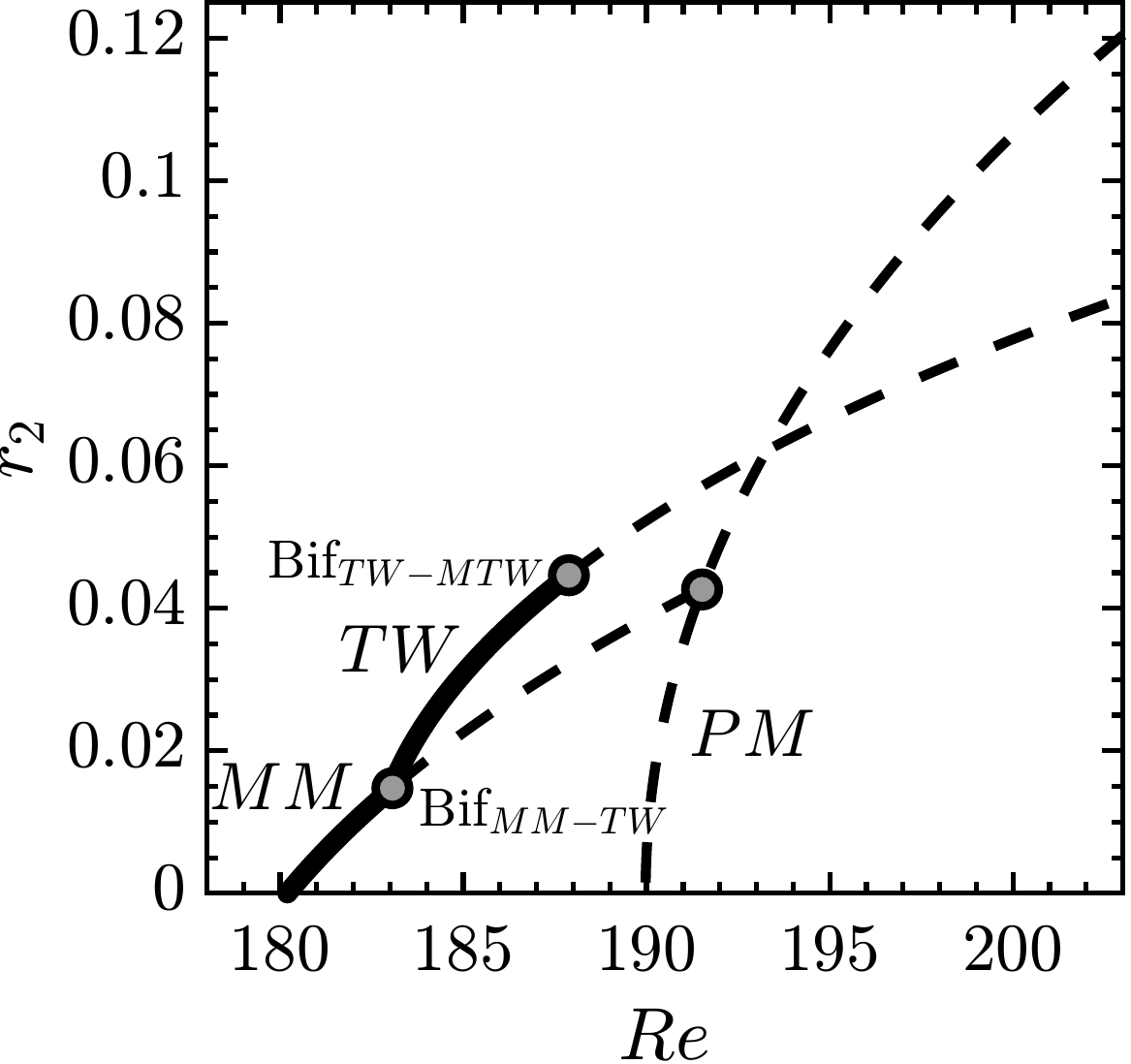} \quad \includegraphics[width=0.55\columnwidth]{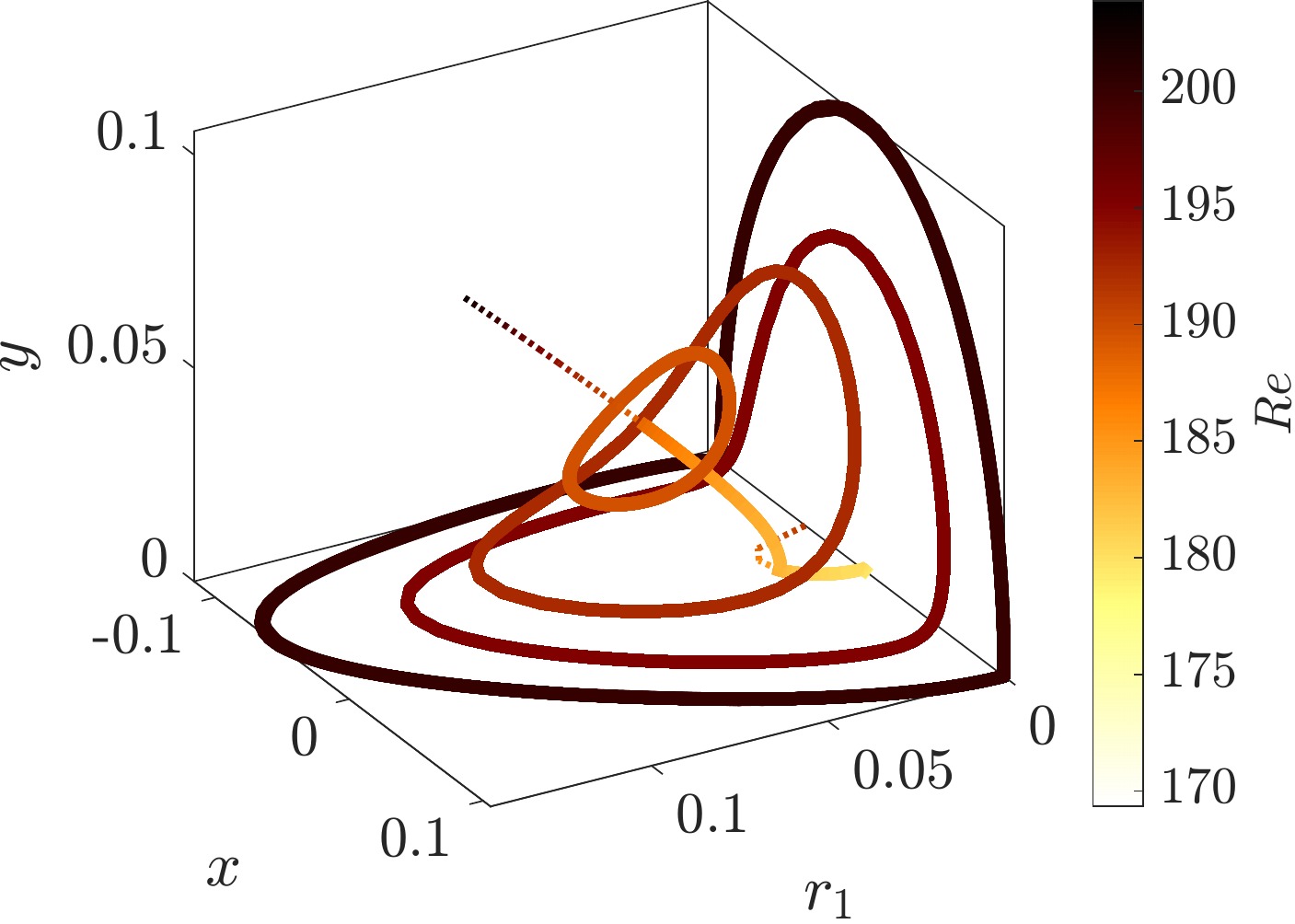}
    \caption{ Bifurcation diagram with respect to the Reynolds number for $L = 1.15$ and $\delta_u = 0.8$. The left diagram reports the evolution of $r_2$ for the fixed point solutions of the normal form. The right diagram displays the bifurcation diagram in the Cartesian coordinates. Solid lines and dashed lines denote stable attractors and unstable attractors, respectively. }
    \label{fig:BifDiagrams}
\end{figure}

We consider the bifurcation sequence for $\delta_u = 1.0$ and $L=1.15$, which is qualitatively similar to transitions in the range $0.5 < \delta_u < 1.5$, near the codimension two points, which are depicted in  \cref{fig:CodimensionTwoEvolution}. At the codimension two points for $\delta_u < 0.5$, at least one of the two bifurcations is sub-critical and a normal form reduction up to fifth order is necessary. Subcritical transition was also noticed for a distance between jets $L=0.1$ by \cite{Cantonetal2017}, who reported high levels of the linear gain associated to transient growth mechanisms.
This last case is out of the scope of the present manuscript.  \Cref{fig:PatternFormationSteadySteady} displays the phase portrait of the stable attractors near the $S_1:S_2$ interaction. For values of $\delta_u > 1.0$, the axisymmetric steady-state loses its axisymmetry leading to a new steady-state with symmetry $m=2$, herein denoted as pure mode (P). A reconstruction of the fluctuating flow field of such a state is performed at the bottom right of \cref{fig:PatternFormationSteadySteady}, which shows that the state P possesses two orthogonal planes of symmetry. Near the codimension two point, for values of the velocity ratio $\delta_u < 1.1$, the state P is only observable, that is non-linearly stable, within a small interval with respect to the Reynolds number. For larger values of the velocity ratio, the state P remains stable within the analysed interval of Reynolds numbers. For values of the velocity ratio  $\delta_u < 1.0$, the bifurcation diagram is more complex. \Cref{fig:BifDiagrams} displays the bifurcation diagram of the fixed-point solutions of \cref{eq:NormalFormThirdOrderSteadySteadyCartesian} on the left diagram and the full set of solutions of the normal form in the right diagram. The axisymmetric steady-state first bifurcates towards a Mixed-Mode solution, which is the solution in the $y=0$ plane for the right diagram of \cref{fig:BifDiagrams}. A solution with a non-symmetric wake has been reconstructed in \cref{fig:PatternFormationSteadySteady}.  The Mixed-Mode solution is only stable within a small interval of the Reynolds number. A secondary bifurcation, denoted Bif$_{MM-TW}$, gives raise to a slowly rotating wave of the wake. The TW and the MM solutions are identical at the bifurcation point. The phase speed is zero at the bifurcation, thus this is not a Hopf bifurcation. It corresponds to a \textit{drift instability} that breaks the azimuthal symmetry, i.e. it starts to slowly drift. 
This unusual feature, that travelling waves bifurcate from a steady solution at a steady bifurcation, is  a generic feature of the $1:2$ resonance. A reconstruction of the travelling wave solution is depicted on the top of \cref{fig:PatternFormationSteadySteady}. It corresponds to the line with non-zero $y$ component in the right diagram of \cref{fig:BifDiagrams}. The TW solution loses its stability in a tertiary bifurcation, denoted as Bif$_{TW-MTW}$. It conforms to a Hopf bifurcation of the TW solution, which gives birth to a quasi-periodic solution name Modulated Travelling Wave (MTW). A representation of this kind of solution in the Cartesian coordinates $(r_1,x,y)$ is depicted on the right image of \cref{fig:BifDiagrams}.    

\begin{figure}
    \centering
    \includegraphics[width=0.9\columnwidth]{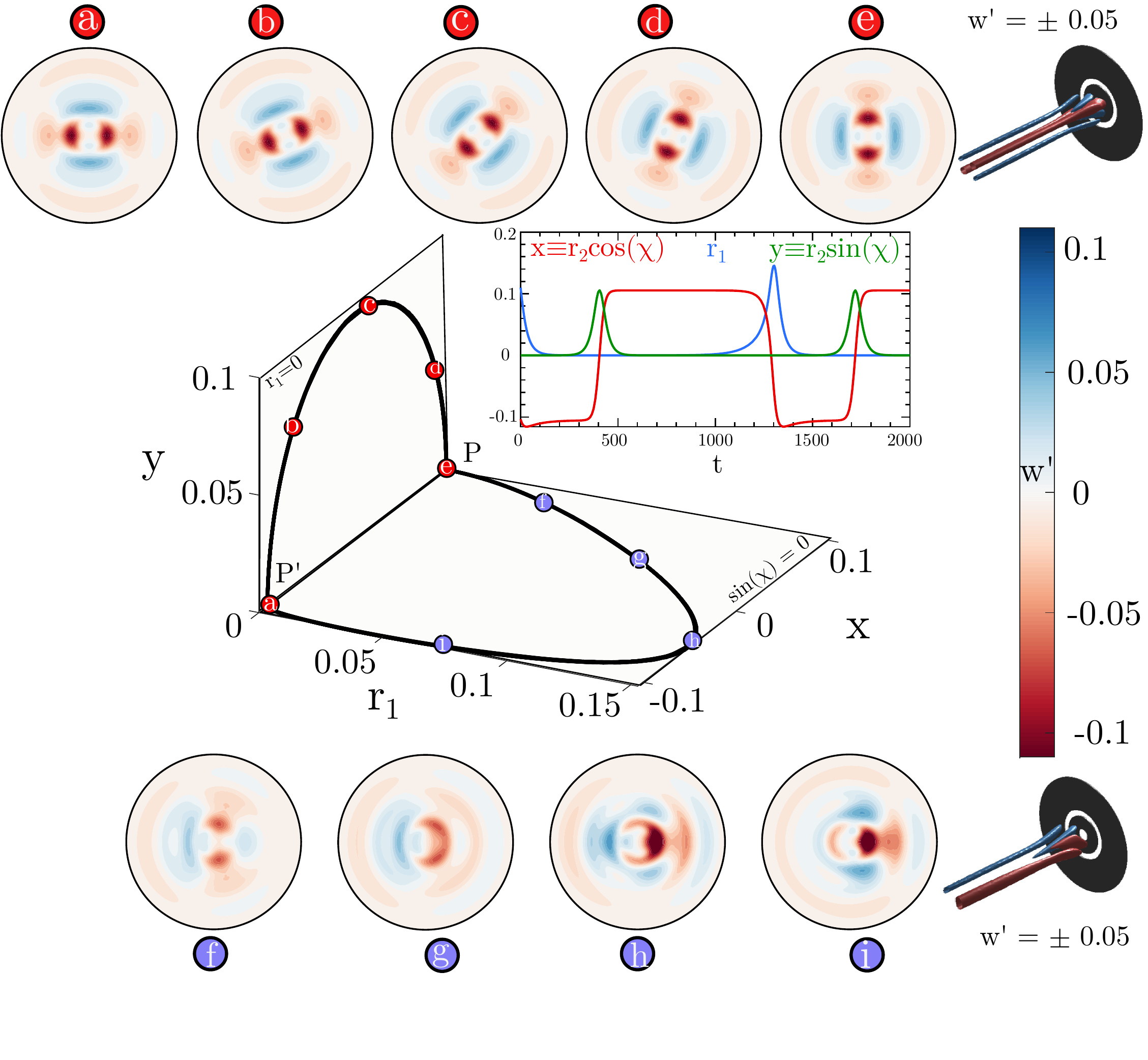} \quad
    \caption{ Heteroclinic cycle solution for parameter values $\text{Re}=200$, $\delta_u = 0.8$. The top and bottom image sequences along the heteroclinic cycle show (from left to right) an axial slice plane at $z=1$ of the instantaneous fluctuations of the axial velocity of the flow field as viewed from downstream, along with a three-dimensional isometric view (d on the top and g on the bottom).The middle diagram displays the heteroclinic cycle in the coordinates ($r_1,x,y$). }
    \label{fig:HeteroclinicCycleAGH}
\end{figure}

Eventually, the Modulated Travelling Wave experiences a global bifurcation. That occurs when the periodic MTW solution, in the $(r_1,x,y)$ coordinates, nearly intersects the invariant $r_1=0$ and $y=0$ planes. The transition sequence is represented in the right image of \cref{fig:BifDiagrams} in the  Cartesian coordinates ($r_1,x,y$). The amplitude of the MTW limit cycle increases until the MTW arising at the tertiary bifurcation Bif$_{TW-MTW}$ are destroyed by meeting a heteroclinic cycle at  Bif$_{MTW-Ht}$. The locus of Bif$_{MTW-Ht}$ is reported in \cref{fig:PatternFormationSteadySteady} and in good agreement with \cite{armbruster1988heteroclinic}. The conditions for the existence of the heteroclinic cycles are listed in \cref{tab:DefinitionModesHt}. When $\sigma_{-}$  becomes negative, the cycle is attracting and robust heteroclinic cycles are observed. It is destroyed when $\sigma_{+}$ becomes negative, in that case the pure modes are no longer saddles which breaks the heteroclinic connection. \Cref{fig:HeteroclinicCycleAGH} displays the instantaneous fluctuation field from a heteroclinic orbit connecting P and its conjugate solution P', which is obtained by a rotation of $\pi/2$, for parameter values $\text{Re} = 200$ and $\delta_u = 0.8$. The dynamics of the cycle takes place in two phases.
\Cref{fig:HeteroclinicCycleAGH} depicts the motion of the coherent structure associated to the heteroclitic cycle. Starting from the conjugated pure mode P', the cycle leaves the point (a), located in the vicinity of P', along the unstable eigenvector $y$, which is the stable direction of P.  
The first phase consists in a rapid rotation by $\pi/2$ of the wake, it corresponds to the sequence a-b-c-d-e displayed in \cref{fig:HeteroclinicCycleAGH}. Then it is followed by a slow approach following the direction $y$ and departure from the pure mode state P along the direction $r_1$. The second phase consists in a rapid horizontal motion of the wake, which is an evolution from P to P' that takes place by the breaking of the reflectional symmetry with respect to the vertical axis; it constitutes the sequence e-f-g-h-i-a. Please note that equivalent motions are also possible. The first phase of rapid counter-clockwise rotation by $\pi/2$ can be performed in the opposite sense. It corresponds to a motion in the Cartesian coordinates along the plane $r_1$ along negative values of $y$. The sequence e-f-g-h-i-a can be replaced by a horizontal movement in the opposite sense, which adjusts to connect the plane $y=0$ corresponding to negative values of $r_1$,

\section{Discussion \& Conclusions \label{sec:conclusions}}
This article achieves a complete description of the configuration consisting of two coaxial jets, broadly found in industrial processes, covering a wide range of applications such as noise reduction, mixing enhancement, or combustion control. The numerical procedure herein employed has been validated with the existing literature in the case of the stability analysis (see \ref{sec:validationCanton} for a detailed overview), and compared to DNS results, as done in \cite{sierra2022Lorite}. The analysis comprises a layout with a wide range of the velocity ratio ($\delta_u=U_i/U_o$) between the jets, from $\delta_u=0$ to $\delta_u=2$, as well as the distance between jets ($L$) enclosing values from $L=0.5$ to $L=4$, substantially expanding the work of \cite{Cantonetal2017}. 

A linear stability analysis reveals the most significant modes, consisting of two steady modes ($S_1$ and $S_2$, located within the recirculation bubble) and two unsteady ones ($F_1$ and $F_2$, evolving as a transient growth in the downstream direction). The critical Reynolds number is determined for a wide range of velocity and distance ratios, starting with the influence of the velocity ratio. As the relation between inner and outer velocities grows, the flow is stabilised, increasing the critical Reynolds number.  The primary instability swaps from mode $S_1$, characterised with one symmetry plane, to mode $S_2$ that possesses two symmetry planes. An abrupt divergence in the critical Reynolds number is captured, associated with the vanishing of the recirculation region, that could suggest a stability control strategy. Subsequently, the effect of the distance $L$ between jets is analysed. The primary effect of increasing this distance is a decrease in the critical Reynolds number for all values of $\delta_u$ investigated. Additionally, the existence of saddle node bifurcations, that swap the most unstable mode of the flow, generates turning points in the neutral curve.

The investigated bifurcation scenario starts from the codimension point, with an axisymmetric steady state located at a velocity ratio $\delta_u = 1.0$ and distance between jets of $L=1.15$. It is qualitatively equivalent to transitions found in the range $0.5<\delta_u<1.5$. It reveals a break of the axisymmetry for values higher than $\delta_u=1.0$, presenting a steady state as a pure mode P with two orthogonal planes of symmetry. For values lower than $\delta_u=1.0$, the bifurcation diagram exhibits a slightly complicated path. Firstly, it drives into a Mixed-Mode (MM) solution presenting a non-symmetric wake, that is only stable for a small range of the Reynolds number. Subsequently, a slowly rotating wake is triggered in the form of a Travelling Wave (TW). This unusual feature, an unsteady state emerging from a steady state, corresponds to a drift instability commonly found at $1:2$ resonance. Then, the TW solution encounters a Hopf bifurcation, developing a quasi-periodic solution in the form of a Modulated Travelling Wave (MTW). Finally, the MTW solution undergoes a global bifurcation meeting a heteroclinic cycle (Ht). This heteroclinic orbit links the solution P with its conjugate solution P', spinning the wake from P' to P, and moving it horizontally from P to P'.
\newpage
\section*{Acknowledgments}
A.C., J.A.M. and S.L.C. acknowledge the grant PID2020-114173RB-I00 funded by MCIN/AEI/ 10.13039/501100011033. S.L.C., J.A.M. and A.C. acknowledge the support of Comunidad de Madrid through the call Research Grants for Young Investigators from Universidad Politécnica de Madrid. AC also acknowledges the support of Universidad Politécnica de Madrid, under the programme ‘Programa Propio’.

\appendix

\section{Normal form reduction}
\label{sec:Appendix_reductionprocedureCoefficients}
In this section we provide a detail explanation of the normal form reduction to obtain the coefficients of \cref{eq:dynsyst}, we define the terms of the compact notation of the governing equations \cref{eq:GoverningEquationsNSCompact}, which is reminded here, for the sake of conciseness, 
 \be{GoverningEquationsNSCompactAppendix}
  \Bop \frac{\partial \QV}{\partial t} = \Fop(\QV, \bm{\eta}) \equiv \Lop \QV  + \Nop(\QV,\QV) + \Gop (\QV,\bm{\eta}).
\ee
The nonlinear convective operator $\vec{N}(\vec{Q}_1,\vec{Q}_2) = \vec{U}_1 \cdot \nabla \vec{U}_2$ accounts for the quadratic interaction on the state variable. The linear operator on the state variable is $\vec{L} \vec{Q} = [\nabla P, \nabla \cdot \vec{U} ]^T$. The remaining term accounts for the linear variations in the state variable and the parameter vector. It is defined as $\vec{G}(\vec{Q},\bm{\eta}) = \vec{G}(\vec{Q},[\eta_1,0]^T) + \vec{G}(\vec{Q},[0,\eta_2]^T)$ where $\vec{G}(\vec{Q},[\eta_1,0]^T) = \eta_1 \nabla \cdot (\nabla \vec{U} + \nabla \vec{U}^T)$ and $\vec{G}(\vec{Q},[0,\eta_2]^T)$. The former operator shows the dependency on the parameter $\eta_1$, which accounts for the viscous effects. The latter operator depends on the parameter $\eta_2$, which  accounts for the velocity ratio between jets and it is used to impose the boundary condition $ \vec{U} = (0, \eta_2 \tanh{ \big(b_i (1 - 2r)\big)} , 0) \text{ on } \Gamma_{in,i}$.
In addition, we consider the following splitting of the parameters $\bm{\eta} = \bm{\eta}_c + \Delta \bm{\eta}$. Here $\bm{\eta}_c$ denotes the critical parameters $\bm{\eta}_c \equiv [Re_c^{-1}, \delta_{u,c}]^T$ attained when the spectra of the Jacobian operator posses at least an eigenvalue whose real part is zero. The distance in the parameter space to the threshold is represented by $\Delta \bm{\eta} = [Re_c^{-1} - Re^{-1}, \delta_{u,c} - \delta_u]^T$.

\subsection{Multiple scales ansatz}
The multiple scales expansion of the solution $\vec{q}$ of \cref{eq:GoverningEquationsNSCompact} is
\be{Normal_Form_Ansatz_WNL}
\displaystyle \vec{q}(t,\tau)  = \QVBF  + \varepsilon \vec{q}_{(\varepsilon)}(t,\tau) + \varepsilon^2  \vec{q}_{(\varepsilon^2)}(t,\tau) + O(\varepsilon^3), 
\ee
where $\varepsilon \ll 1$ is a small parameter. The distance in the parameter space to the critical point $\Delta \bm{\eta} = [Re_c^{-1} - Re^{-1}, \delta_{u,c} - \delta_u]^T$ is assumed to be of second order, i.e. $ \Delta {\eta}_i = O(\varepsilon^2) $ for $i=1,2$. The expansion \cref{eq:Normal_Form_Ansatz_WNL} considers a two scale expansion of the original time $t \mapsto t + \varepsilon^2 \tau$. A fast timescale $t$ and a slow timescale of the evolution of the amplitudes $z_i(\tau)$ in \cref{eq:Normal_Form_Ansatz_WNL}, for $i=1,2$.
Note that the expansion of the LHS \cref{eq:GoverningEquationsNSCompact} up to third order is as follows 
\be{GoverningEqExpandedLHS}
\displaystyle  \varepsilon \vec{B} \frac{\partial \vec{q}_{(\varepsilon)}} {\partial t} + \varepsilon^2 \vec{B} \frac{\partial \vec{q}_{(\varepsilon^2)}} {\partial t} 
+ \varepsilon^3 \big[ \vec{B} \frac{\partial \vec{q}_{(\varepsilon^3)}} {\partial t}  + \vec{B} \frac{\partial \vec{q}_{(\varepsilon)}} {\partial \tau} \big],
\ee
and the RHS respectively,
\be{GoverningEqExpandedRHS}
\vec{F} (\vec{q}, \bm{\eta}) = \vec{F}_{(0)} + \varepsilon \vec{F}_{(\varepsilon)}  + \varepsilon^2 \vec{F}_{(\varepsilon^2)}  + \varepsilon^3 \vec{F}_{(\varepsilon^3)}.
\ee
The expansion \cref{eq:GoverningEqExpandedRHS} will be detailed at each order. 

\subsubsection{Order $\varepsilon^0$}
The zeroth order $\QVBF$ of the multiple scales expansion \cref{eq:Normal_Form_Ansatz_WNL} is the steady state of the governing equations evaluated at the threshold of instability, i.e. $\bm{\eta} = \bm{\eta}_c$,

\be{ZerothOrder}
\displaystyle \vec{0} = \vec{F} (\QVBF, \bm{\eta}_c).
\ee

\subsubsection{Order $\varepsilon^1$}
The first order $\vec{q}_{(\varepsilon)}(t,\tau)$ of the multiple scales expansion of \cref{eq:Normal_Form_Ansatz_WNL} is composed of the eigenmodes of the linearized system
\be{FirstOrderA}
    \vec{q}_{(\varepsilon)}(t,\tau) \equiv  \big(  z_1(\tau)  e^{-\im m_1 \theta} \hat{\vec{q}}_{1} +  z_2(\tau)  e^{\im -m_2 \theta} \hat{\vec{q}}_{2}  + \text{c. c.} \big).
\ee
in our case, $m_1 = 1$ and $m_2 = 2$.
Each term $\hat{\vec{q}}_\ell$ of the first order expansion \cref{eq:FirstOrderA} is a solution of the following linear equation
\be{FirstOrderB}
 \vec{J}_{(\omega_\ell, m_\ell)} \hat{\vec{q}}_{\ell} = \Big(i\omega_\ell \vec{B} - \frac{\partial \vec{F}}{\partial \vec{q} }|_{\vec{q} = \QVBF, \bm{\eta} = \bm{\eta}_c} \Big) \hat{\vec{q}}_{\ell},
\ee
where $\frac{\partial \vec{F}}{\partial \vec{q} }|_{\vec{q} = \QVBF, \bm{\eta} = \bm{\eta}_c} \hat{\vec{q}}_{\ell} = \vec{L}_{m_\ell} \hat{\vec{q}}_{\ell} + \vec{N}_{m_\ell}(\QVBF, \hat{\vec{q}}_{\ell} ) + \vec{N}_{m_\ell}(\hat{\vec{q}}_{\ell}, \QVBF )$. The subscript $m_\ell$ indicates the azimuthal wavenumber used for the evaluation of the operator. 

\subsubsection{Order $\varepsilon^2$}
The second order expansion term $\vec{q}_{(\varepsilon^2)}(t,\tau)$ is determined from the resolution of a set of forced linear systems, where the forcing terms are evaluated from first and zeroth order terms. The expansion in terms of amplitudes $z_i(\tau)$ ($i=1,2$) of $\vec{q}_{(\varepsilon^2)}(t,\tau)$ is assessed from term-by-term identification of the forcing terms at the second order. Non-linear second order terms in $\varepsilon$ are
\bes{SecondOrderA}
\displaystyle \vec{F}_{(\varepsilon^2)} & \equiv \displaystyle  \sum_{j,k=1}^2 \Big( z_j z_k \vec{N}(\hat{\vec{q}}_j,\hat{\vec{q}}_k) e^{-\im(m_j+m_k)\theta} + \text{ c.c.} \Big) \\
& + \displaystyle \sum_{j,k=1}^2 \Big( z_j \overline{z}_k \vec{N}(\hat{\vec{q}}_j,\overline{\hat{\vec{q}}}_k) e^{-\im(m_j-m_k)\theta} + \text{ c.c.} \Big) \\ 
 & + \displaystyle \sum_{\ell=0}^2 \eta_\ell \vec{G}(\QVBF, \vec{e}_\ell),
\ees
where the terms proportional to $z_j z_k$ are named $\hat{\vec{F}}_{(\epsilon^2)}^{(z_j z_k)}$ and  $\vec{e}_\ell$ is an element of the orthonormal basis of $\mathbb{R}^2$.

Then, we look for a second order term expanded as follows
\be{SecondOrderB}
 \vec{q}_{(\varepsilon^2)} \equiv \displaystyle \sum_{\substack{j,k=1 \\ k\leq j}}^{2} \big( z_j z_k \hat{\vec{q}}_{z_j z_k} + z_j \overline{z}_k \hat{\vec{q}}_{z_j \overline{z}_k} + \text{ c.c } \big) + \displaystyle  \sum_{\ell=1}^2 \eta_\ell  \QVBF^{(\eta_\ell)}.
\ee
 Terms $\hat{\vec{q}}_{z^2_j}$ are azimuthal harmonics of the flow. The terms $\hat{\vec{q}}_{z_j z_k}$ with $j \neq k$ are coupling terms, and  $\hat{\vec{q}}_{|z_j|^2}$ are harmonic base flow modification terms. Finally, $\QVBF^{(\eta_\ell)}$ are base flow corrections due to a variation of the parameter $\eta_\ell$ from the critical point. 

At this order, there exists two resonant terms, the terms proportional to $\overline{z}_1 z_2$ and $z_1^2$, which are associated with the singular Jacobian $\vec{J}_{(0,m_k)}$ for $k=1,2$. To ensure the solvability of these terms, we must enforce compatibility conditions, i.e. the \textit{Fredholm alternative}. The resonant terms are then determined from the resolution of the following set of \textit{bordered systems}
\begin{equation}
\begin{pmatrix} \vec{J}_{(0,m_k)} & \hat{\vec{q}}_{k} \\ \hat{\vec{q}}_{k}^\dagger & 0 \end{pmatrix} \begin{pmatrix} \hat{\vec{q}}_{(\vec{z}^{(R)})} \\ e \end{pmatrix} = \left( \begin{array}{c} \hat{\vec{F}}^{(\vec{z}^{(R)})}_{(\varepsilon^2)} \\ 0 \end{array} \right), \text{ } \vec{z}^{(R)} \in [\overline{z}_1 z_2, z_1^2]^T,
\end{equation}
where $e=e_3$ for $\vec{z}^{(R)} = \overline{z}_1 z_2$ and $e=e_4$ for $\vec{z}^{(R)} =  z_1^2$. The non-resonant terms are computed by solving the following non-degenerated forced linear systems
\be{SecondOrderC}
\displaystyle \vec{J}_{(0, m_j + m_k)} \hat{\vec{q}}_{z_j z_k} = \hat{\vec{F}}_{(\epsilon^2)}^{(z_j z_k)},
\ee
and 
\be{SecondOrderD}
\displaystyle \vec{J}_{(0,0)} \QVBF^{(\eta_\ell)} = \vec{G}(\QVBF, \vec{e}_\ell).
\ee

\subsubsection{Order $\varepsilon^3$}

At third order, there exists six degenerate terms. In our case, we are not interested in solving for terms of third-order, instead, we will determine the linear and cubic coefficients of the third order normal form \cref{eq:dynsyst} from a set of compatibility conditions. 

The linear terms $\lambda_{(s,1)}$ and $\lambda_{s,2}$ and cubic terms $c_{(i,j)}$ for $i=1,2$ are determined as follows
\be{ThirdOrderA}
\lambda_{(s,1)} = \frac{\langle \hat{\vec{q}}_1^\dagger, \hat{\vec{F}}^{(z_1)}_{(\varepsilon^3)} \rangle}{\langle \hat{\vec{q}}_z^\dagger, \vec{B}  \hat{\vec{q}}_z \rangle }, \text{ } \lambda_{(s,2)} = \frac{\langle \hat{\vec{q}}_2^\dagger, \hat{\vec{F}}^{(z_2)}_{(\varepsilon^3)} \rangle}{\langle \hat{\vec{q}}_2^\dagger, \vec{B}  \hat{\vec{q}}_2 \rangle }, \text{ } c_{(i,j)} = \frac{\langle \hat{\vec{q}}_2^\dagger, \hat{\vec{F}}^{(z_i |z_j|^2)}_{(\varepsilon^3)} \rangle}{\langle \hat{\vec{q}}_i^\dagger, \vec{B}  \hat{\vec{q}}_i \rangle }.
\ee

The forcing terms for the linear coefficient are 
\be{ForcingTerm_ThirdOrder_Linear}
\hat{\vec{F}}^{(z_j)}_{(\varepsilon^3)} \equiv \sum_{\ell=1}^2 \eta_\ell \Big( \big[ \vec{N}(\hat{\vec{q}}_j,\QVBF^{(\eta_\ell))} + \vec{N}(\QVBF^{(\eta_\ell)},\hat{\vec{q}}_j) \big] + \vec{G}(\hat{\vec{q}}_j, \vec{e}_\ell) \Big).
\ee
which allows the decomposition of $\lambda_{(s,\ell)} = \lambda_{(s,\ell),\text{Re}} (\text{Re}^{-1}_c \text{Re}^{-1}) + \lambda_{(s,\ell),{\delta_u}} (\delta_{u,c}- \delta_u)$ for $\ell = 1,2$.

The forcing terms for the cubic coefficients are
\bes{ForcingTerm_ThirdOrder_CoeffA}
\hat{\vec{F}}^{({z}_j |z_k|^2)}_{(\varepsilon^3)} & \equiv \big[ \vec{N}(\hat{\vec{q}}_j,\hat{\vec{q}}_{|z_k|^2}) + \vec{N}(\hat{\vec{q}}_{|z_k|^2},\hat{\vec{q}}_j) \big] \\
& + \big[ \vec{N}(\hat{\vec{q}}_{-k},\hat{\vec{q}}_{z_j z_k}) + \vec{N}(\hat{\vec{q}}_{j,k},\hat{\vec{q}}_{-k}) \big] \\
&  + \big[ \vec{N}(\hat{\vec{q}}_{k},\hat{\vec{q}}_{z_j \overline{z}_k}) + \vec{N}(\hat{\vec{q}}_{z_j \overline{z}_k},\hat{\vec{q}}_{k}) \big].
\ees
if $j \neq k$
and 
\bes{ForcingTerm_ThirdOrder_CoeffB}
\hat{\vec{F}}^{({z}_j |z_j|^2)}_{(\epsilon^3)} & \equiv \big[ \vec{N}(\hat{\vec{q}}_j,\hat{\vec{q}}_{|z|_j^2}) + \vec{N}(\hat{\vec{q}}_{|z|_j^2},\hat{\vec{q}}_j) \big] \\
& + \big[ \vec{N}(\hat{\vec{q}}_{-j},\hat{\vec{q}}_{z_j^2}) + \vec{N}(\hat{\vec{q}}_{z^2_j},\hat{\vec{q}}_{-j}) \big],
\ees
for the diagonal forcing terms.

\section{Validation of the code - Comparison with the literature}
\label{sec:validationCanton}
The calculations made in StabFem in the sections at the main manuscript are validated comparing the leading global mode in the geometry proposed by \cite{Cantonetal2017}. Moreover, the critical Reynolds number and the frequency associated are also analysed. In the cited work, the authors use an analogous geometry with the following parameters:

\begin{itemize}
    \item Radious of the inner jet $R_{inner} = 0.5$
    \item Diameter of the outer jet $D = 0.4$
    \item Distance between jets $L = 0.1$
    \item Ratio between velocities $\delta_u = 1$ 
\end{itemize}

The linear stability analysis has been carried out imposing $m = 0$, as done by \cite{Cantonetal2017}, so the leading global mode will be axisymmetric. The critical Reynolds number $Re_c$ and the frequency $\omega$ of the leading global mode are compared in  Tab. \ref{tab:comparison}. As seen, few differences can be found on the critical Reynolds number and the frequency. The relative error in the $Re_c$ calculation is $1.06\%$ and the one of the frequency is $0.17 \%$.

\begin{table}
    \centering
    \begin{tabular}{c|c|c}
         & Canton \textit{et al.} (2017) \cite{Cantonetal2017} & Present work \\
        $Re_c$ & $1420$ & $1405$\\
        $\omega$& $5.73$ & $5.72$\\
    \end{tabular}
    \caption{Comparison of $Re_c$ and $\omega$ between previous work and the present one.}
    \label{tab:comparison}
\end{table}

The global mode is now calculated using StabFem and compared with the one calculated by \cite{Cantonetal2017}. This mode can be found in figures 9, 10 and 11 on the cited paper. As it can be seen, there are not substantial differences between the direct modes, being both of them a vortex street with their biggest amplitude situated 10 units downstream the exit of the jets. The adjoint mode is concentrated within the nozzle, with its biggest amplitude situated on the sharp corners. There is not any difference between the adjoint mode calculated with StabFem and the one in \cite{Cantonetal2017}. Finally, the structural sensitivity is similar to the one computed by \cite{Cantonetal2017}. It is composed by two lobes in the space between the exit of the two jets.

\begin{figure}
    \centering
    \includegraphics[width=0.9\columnwidth]{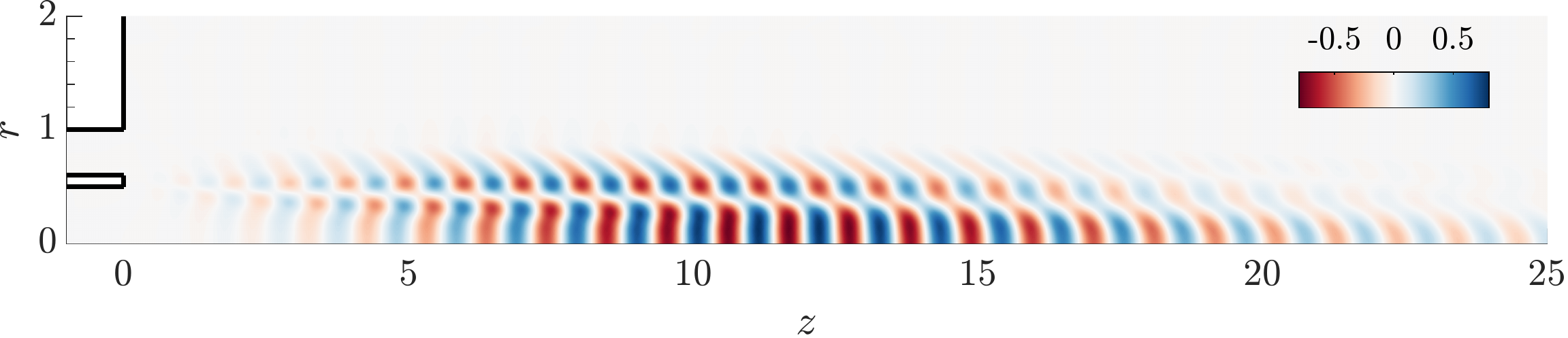}
    \includegraphics[height=0.25\columnwidth]{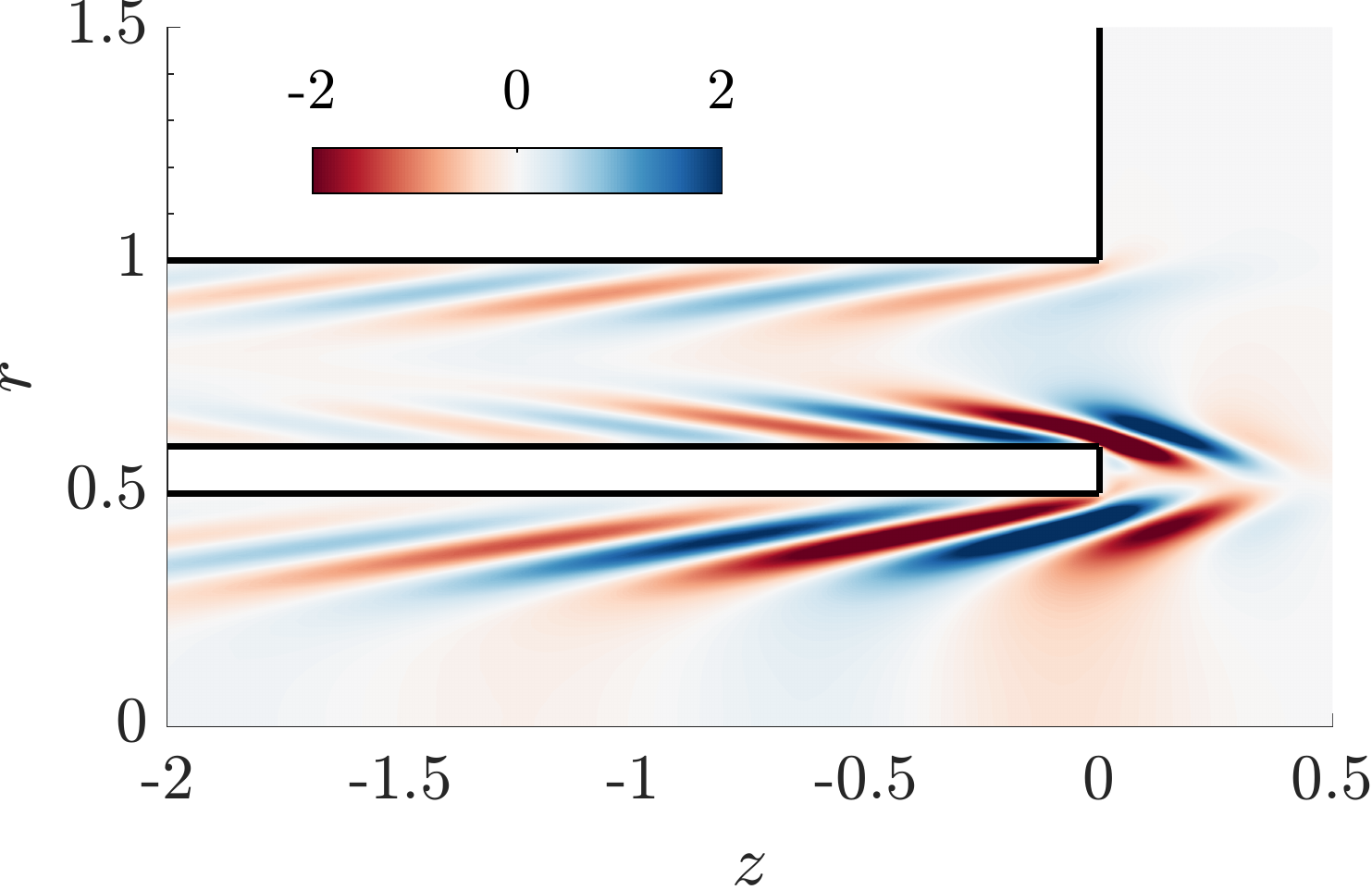}
    \includegraphics[height=0.25\columnwidth]{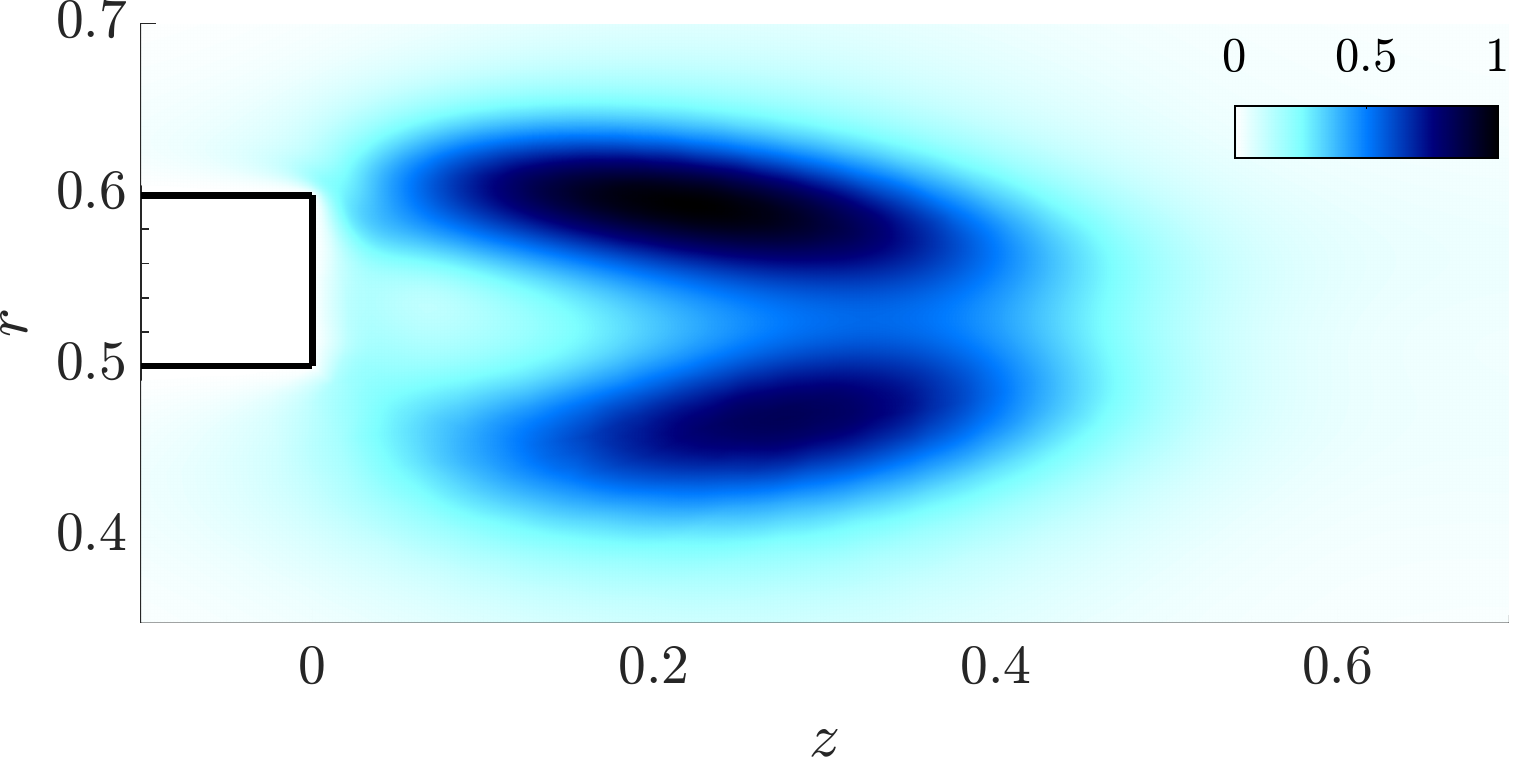}
    \caption{Direct mode, adjoint mode and sensitivity of the leading global mode studied by \cite{Cantonetal2017} calculated using StabFem.}
    \label{fig:mode_validation}
\end{figure}

\bibliography{jfmBib}

\end{document}